\documentclass[fleqn,twoside]{article}
\usepackage{espcrc2}

\newcommand{\AmS}{{\protect\the\textfont2
  A\kern-.1667em\lower.5ex\hbox{M}\kern-.125emS}}

\title{Towards Configuration of applied Web-based information system}

\author{Mark Sh. Levin
\thanks{
 Mark Sh. Levin:~
 Inst. for Information Transmission Problems,
 Russian Academy of Sciences, 19 Bolshoj Karetny lane, Moscow 127994, Russia.
 Email: mslevin@acm.org
  }
  }

\begin{document}

\begin{abstract}
In the paper, combinatorial synthesis of structure for
 applied
 Web-based
 systems is described.
 The problem is considered as a combination
 of selected design alternatives for system parts/components
 into a resultant composite decision (i.e., system configuration design).
  The solving framework is based on Hierarchical Morphological
 Multicriteria Design (HMMD) approach:
 (i) multicriteria selection of alternatives for system
 parts,
 (ii) composing the selected alternatives
 into a resultant combination
 (while taking into
 account ordinal quality of the alternatives above
 and their compatibility).
 A lattice-based discrete space is used to evaluate
 (to integrate)
 quality of the resultant
 combinations
 (i.e., composite system decisions or system configurations).
 In addition, a simplified solving framework based on multicriteria
 multiple choice problem
 is considered.
 A multistage design process to obtain a system trajectory
 is described as well.
 The basic applied example is targeted to
 an applied Web-based system for a communication service
 provider.
 Two other applications are briefly described
 (corporate system and
 information system for academic
 application).

~~


\vspace{1pc}
\end{abstract}

\maketitle

\newcounter{cms}
\setlength{\unitlength}{1mm}

\section{Introduction}

%
 Web-based
 applied
 systems
 are
 increasing in popularity.
 Here some basic technological directions can be pointed out
 as follows:

 (a) E-business and E-commerce, for example:
 (i) smart marketplaces  are presented in
   \cite{geng03},
  (ii) Web services are studied in
  (\cite{leymann02}
   \cite{younas05}),
  (iii) E-commerce books are presented
 in (\cite{rey04}
%
     \cite{shaw00});

%
 (b) Web-based information systems are
 studied in
  (\cite{feng06},
  \cite{isa98},
   \cite{taka97});

%
 (c) E-government and E-democracy,
  for example:
  (i) e-government systems  (including their management,
  functionality, evolution, designing, and innovation)
   are presented in
    (\cite{heeks05},
      \cite{layne01},
     \cite{moon02},
   \cite{prins06},
    \cite{scherlis03});
  (ii) decision support for participatory democracy
 is presented in
  \cite{insua08},
  (iii) Web-based public participation geographical information systems
 are described in
   \cite{kingston00};

%
 (d) Web-based medicine systems:
 (i) Web-based telemedicine systems for home-care
 are presented in
  \cite{bellazzi01},
 (ii) development of Web-based clinical information systems
 was introduced in
  \cite{cimino95});

%
 (e) Web-based educational systems
 (e-learning, e-teaching, etc.), for example:
 (i) Web-based learning and teaching technologies are studied
 in
 \cite{aggar00},
 (ii) development of adaptive Web-based courseware
   was introduced in
  \cite{brus98},
  (iii) building a Web-based educational system
  was presented in
  \cite{mccormack97},
  (iv) conceptual view of web-based e-learning was suggested in
  \cite{schewe05};

 (f) Web-based research support systems, for example:
 (i) special Web-based research support system was
     designed
   \cite{tang03},
 (ii) framework for Web-based research support was suggested in
  \cite{yao03}.
%
%
%
%


 %
 Generally,
 it is possible to consider the following brief description of
 an applied Web-based
 system:
 (1) there is a set of users and server(s),
 (2) each user has information and computing tasks
 (including Web-based tasks),
 (3) the server is a basis for information system
 (i.e., information processing)
  and computing
   (for each user),
 (4) each user has a personal browser,
 (5) users have their access to the server(s) separately,
 (6) there is a concurrent multiple user access,
 (7) there are limitations to the volume of information transmission,
 and
 (8) there are
  requirements to performance, security, scalability, adaptability
  and upgradeability.
%
%
%
%
%
%
%
%
 As a result,
 there exists a need of
 Web-based system life cycle engineering/management
 (e.g., Web engineering)
  including
 requirements engineering, design, maintenance
 (e.g.,
  \cite{bat01},
  \cite{benat02},
 \cite{feng06},
 \cite{fra99},
 \cite{geller97},
 \cite{ginige01},
 \cite{huang05},
 \cite{kappel06},
 \cite{land07},
 \cite{leymann02},
 \cite{lin08},
 \cite{mano05},
 \cite{mccormack97},
 \cite{papa06},
 \cite{rey04},
 \cite{yu05},
  \cite{younas05}).


%
%
 Mainly, the design process of
 Web-based
 applied
  systems
  consists in
 system configuration design
 (i.e., selection or composition of
 design alternatives for
  system components/parts)
 (e.g.,
 \cite{arpinar05},
 \cite{benat02},
 \cite{benat03},
 \cite{blake05},
 \cite{dust05},
  \cite{geller97},
  \cite{huang05},
 \cite{mad06},
 \cite{milanovic04},
 \cite{oh05},
 \cite{sung08},
  \cite{yu05}).
%
 Fig. 1 illustrates the design of system configuration as a
 selection of alternatives for system parts.
  Here a composite (modular)
 system consists of \(m\) system parts:
 \( \{  P(1), ... ,P(i), ... ,P(m)  \} \).
 For each system part (i.e., \(\forall i, ~i=\overline{1,m}\))
 there are corresponding alternatives
 \( \{ X^{i}_{1}, X^{i}_{2}, ... , X^{i}_{q_{i}} \} \), where
 \(q_{i}\) is the number of alternatives for part \(i\).
 The problem is:

  {\it Select an alternative for each system part
 while taking into account some
 local and/or global
 objectives/preferences and constraints.}

\begin{center}
\begin{picture}(70,43)

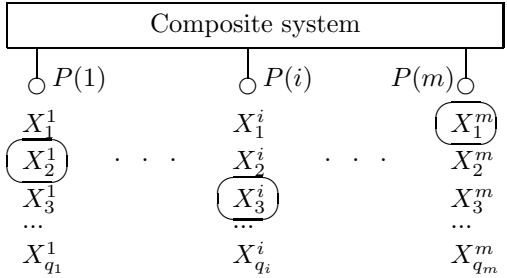
\put(06,00){\makebox(0,0)[bl] {Fig. 1. System configuration
problem}}

\put(4,5){\makebox(0,8)[bl]{\(X^{1}_{q_{1}}\)}}
\put(4,11){\makebox(0,8)[bl]{\(. . .\)}}
\put(4,13){\makebox(0,8)[bl]{\(X^{1}_{3}\)}}
\put(4,18){\makebox(0,8)[bl]{\(X^{1}_{2}\)}}
\put(4,23){\makebox(0,8)[bl]{\(X^{1}_{1}\)}}
\put(06,20){\oval(8,5.6)}

\put(32,5){\makebox(0,8)[bl]{\(X^{i}_{q_{i}}\)}}
\put(32,11){\makebox(0,8)[bl]{\(. . .\)}}
\put(32,13){\makebox(0,8)[bl]{\(X^{i}_{3}\)}}
\put(32,18){\makebox(0,8)[bl]{\(X^{i}_{2}\)}}
\put(32,23){\makebox(0,8)[bl]{\(X^{i}_{1}\)}}
\put(34,15){\oval(8,5.6)}

\put(61,5){\makebox(0,8)[bl]{\(X^{m}_{q_{m}}\)}}
\put(61,11){\makebox(0,8)[bl]{\(. . .\)}}
\put(61,13){\makebox(0,8)[bl]{\(X^{m}_{3}\)}}
\put(61,18){\makebox(0,8)[bl]{\(X^{m}_{2}\)}}
\put(61,23){\makebox(0,8)[bl]{\(X^{m}_{1}\)}}
\put(63,25){\oval(8,5.6)}

\put(6,30){\circle{2}} \put(34,30){\circle{2}}
\put(63,30){\circle{2}}

\put(06,31){\line(0,1){4}} \put(34,31){\line(0,1){4}}
\put(63,31){\line(0,1){4}}

\put(08,29){\makebox(0,8)[bl]{\(P(1)\) }}
\put(36,29){\makebox(0,8)[bl]{\(P(i)\) }}
\put(53,29){\makebox(0,8)[bl]{\(P(m)\) }}


\put(15,20){\makebox(0,8)[bl]{ .~ .~ .}}
\put(43,20){\makebox(0,8)[bl]{ .~ .~ .}}


\put(02,35){\line(1,0){66}} \put(02,41){\line(1,0){66}}
\put(02,35){\line(0,1){6}} \put(68,35){\line(0,1){6}}


\put(21,36.5){\makebox(0,8)[bl]{Composite system}}




\end{picture}
\end{center}

 In Fig. 1 the following system configuration example is depicted:
 ~\(S_{1}=X^{1}_{2}\star ... \star X^{i}_{3} \star ... \star X^{m}_{1}\).
 Table 1 contains
 some approaches to composition of
    applied Web-based systems.
%
%
%
%
%
%
%
%
%
%
%
%
%
  Note a survey of combinatorial optimization models, which can be used
  for system configuration design problems,
 is presented in
 \cite{lev09}.

%
%
%


%
%

%
 The paper addresses the problem of combining
 some design alternatives (DAs)
 for Web-based
 system parts/components
 into a resultant composite decision (i.e., system configuration).
 A special lattice-based discrete space is used to evaluate
 quality of the resultant composite system decisions or system configurations.
 The above-mentioned discrete space
  integrates
  ordinal quality of system elements and
 ordinal quality of compatibility among the system elements.

\begin{center}
\begin{picture}(70,79)

\put(0.5,75){\makebox(0,0)[bl] {Table 1. Composition of Web-based
systems}}


\put(00,00){\line(1,0){70}} \put(00,67){\line(1,0){70}}
\put(00,73){\line(1,0){70}}

\put(00,00){\line(0,1){73}} \put(52,00){\line(0,1){73}}
\put(70,00){\line(0,1){73}}


\put(02,68.5){\makebox(0,8)[bl]{Approach }}
\put(53,69){\makebox(0,8)[bl]{References}}


\put(02,62){\makebox(0,8)[bl]{1. Object-oriented approach }}
\put(53,62){\makebox(0,8)[bl]{\cite{geller97}, \cite{rein02}}}


\put(02,58){\makebox(0,8)[bl]{2. Declarative approach }}
\put(53,58){\makebox(0,8)[bl]{\cite{mad06}}}


\put(02,54){\makebox(0,8)[bl]{3. Model-driven design }}
\put(53,54){\makebox(0,8)[bl]{\cite{mano05}}}


\put(02,50){\makebox(0,8)[bl]{4. AI techniques }}
\put(53,50){\makebox(0,8)[bl]{\cite{oh05}}}


\put(02,46){\makebox(0,8)[bl]{5. Self-serv environment }}
\put(53,46){\makebox(0,8)[bl]{\cite{benat03}}}


\put(02,42){\makebox(0,8)[bl]{6. Agent-based approach }}
\put(53,42){\makebox(0,8)[bl]{\cite{blake05}}}


\put(02,38){\makebox(0,8)[bl]{7. Ontology-based approach}}
\put(53,38){\makebox(0,8)[bl]{\cite{arpinar05}}}

\put(02,34){\makebox(0,8)[bl]{8. QoS-aware selection of web}}

\put(06,30){\makebox(0,8)[bl]{services }}

\put(53,34){\makebox(0,8)[bl]{\cite{lin08}}}

\put(02,26){\makebox(0,8)[bl]{9. Dynamic selection }}
\put(53,26){\makebox(0,8)[bl]{\cite{huang05}}}


\put(01,22){\makebox(0,8)[bl]{10. QoS capable Web service }}

\put(06,18){\makebox(0,8)[bl]{composition (multiple choice}}

\put(06,14){\makebox(0,8)[bl]{knapsack problem, )}}

\put(06,10){\makebox(0,8)[bl]{shortest path problem)}}

\put(53,22){\makebox(0,8)[bl]{\cite{yu05}
 }}


\put(01,06){\makebox(0,8)[bl]{11. Petri net approach to com-}}

\put(06,02){\makebox(0,8)[bl]{position of Web services}}

\put(53,06){\makebox(0,8)[bl]{\cite{xiong10}}}

\end{picture}
\end{center}

%
 Hierarchical Morphological
 Multicriteria Design (HMMD) approach
 was suggested by Levin
 (e.g., \cite{lev98},
 \cite{lev05},
  \cite{lev06}).
 This approach is used here as a general solving framework.
 In addition, a simplified solving framework based on usage of multicriteria
 multiple choice problem
 is considered.
 This approach was suggested in
 (\cite{levs06}
  \cite{poladian06}).
  Here the design problem does not involve element compatibility.
 Further a multistage design process to obtain a system trajectory
 is described.
 This design problem was presented in
  (e.g., \cite{lev05}, \cite{lev06}).
 For this problem the solving scheme is based on HMMD.
%

%
 HMMD approach involves the following phases:
 (i) design of a
 system
  tree-like model for the resultant composite decisions,
%
 (ii) generation of (searching for) design alternatives for leaf
 nodes of the system model,
 (iii) evaluation of the alternatives for system parts, and
 (iv) composing the alternatives (DAs)
 into a resultant combination as the system decision(s)
 (while taking into
 account ordinal quality of the alternatives above
 and their compatibility or interconnection).
 Note HMMD generalizes morphological analysis
 that was created by Zwicky
 (e.g., \cite{zwi69}).

 HMMD implements modular multi-stage design approach and provides the following:
 (1) hierarchical (Bottom-Up) design process
 (multicriteria assessment,
 evaluation, selection,
 composition of design alternatives),
 (2) independent assessment and analysis of design alternatives
 for each system part/component
 (including joint and/or independent participation of different domain experts),
 (3) integration of analytical, computer-based, and expert-based
 assessment of design alternatives and their interconnection,
 (4) parallel (and concurrent) analysis and design (evaluation,
 selection, composition) of design alternatives for composite system
 parts/components,
 (5) opportunity to use cognitive methods at each step and/or part of the
 design process.

 In the article,
 the basic applied example is targeted
 to an applied information system for a communication service provider.
 Two other applications are briefly described:
 corporate information system and
 information system for an academic (scientific and/or educational) application.
 The same hierarchical design approach has been used to Web-hosting systems
 \cite{levshar09}.
 Fig. 2 illustrates the introduction part.

\begin{center}
\begin{picture}(73.5,74)

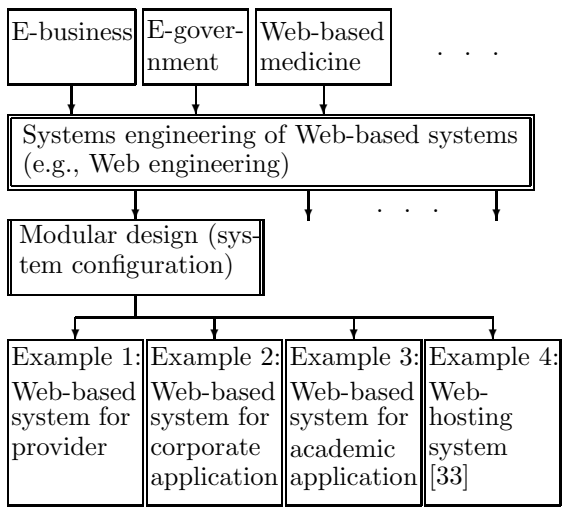
\put(05,00){\makebox(0,0)[bl] {Fig. 2. Illustration for
introduction part}}


\put(00,62){\line(1,0){17}} \put(00,72){\line(1,0){17}}
\put(00,62){\line(0,1){10}} \put(17,62){\line(0,1){10}}

\put(0.5,68){\makebox(0,8)[bl]{E-business}}


\put(8.5,62){\vector(0,-1){4}}


\put(18,62){\line(1,0){14}} \put(18,72){\line(1,0){14}}
\put(18,62){\line(0,1){10}} \put(32,62){\line(0,1){10}}

\put(18.5,67.5){\makebox(0,8)[bl]{E-gover-}}

\put(18.5,64){\makebox(0,8)[bl]{nment}}

\put(25,62){\vector(0,-1){4}}


\put(33,62){\line(1,0){18}} \put(33,72){\line(1,0){18}}
\put(33,62){\line(0,1){10}} \put(51,62){\line(0,1){10}}

\put(33.5,68){\makebox(0,8)[bl]{Web-based}}
\put(33.5,64){\makebox(0,8)[bl]{medicine}}

\put(42,62){\vector(0,-1){4}}






\put(57,66){\makebox(0,8)[bl]{.~ .~ .}}


\put(00,48){\line(1,0){70}} \put(00,58){\line(1,0){70}}
\put(00,48){\line(0,1){10}} \put(70,48){\line(0,1){10}}

\put(00.5,48.5){\line(1,0){69}} \put(00.5,57.5){\line(1,0){69}}
\put(00.5,48.5){\line(0,1){09}} \put(69.5,48.5){\line(0,1){09}}

\put(2,53.5){\makebox(0,8)[bl]{Systems engineering of Web-based
systems}}

\put(2,49.5){\makebox(0,8)[bl]{(e.g., Web engineering)}}

\put(17,48){\vector(0,-1){4}}

\put(40,48){\vector(0,-1){4}}

\put(49,45){\makebox(0,8)[bl]{.~ .~ .}}

\put(65,48){\vector(0,-1){4}}


\put(00,34){\line(1,0){34}} \put(00,44){\line(1,0){34}}
\put(00,34){\line(0,1){10}} \put(34,34){\line(0,1){10}}
\put(00.5,34){\line(0,1){10}} \put(33.5,34){\line(0,1){10}}

\put(1.5,40){\makebox(0,8)[bl]{Modular design (sys-}}
\put(1.5,36){\makebox(0,8)[bl]{tem configuration)}}

\put(17,34){\line(0,-1){3}}

\put(9,31){\line(1,0){55.5}}

\put(9,31){\vector(0,-1){3}} \put(27.5,31){\vector(0,-1){3}}
\put(46,31){\vector(0,-1){3}} \put(64.5,31){\vector(0,-1){3}}


\put(00,06){\line(1,0){18}} \put(00,28){\line(1,0){18}}
\put(00,06){\line(0,1){22}} \put(18,06){\line(0,1){22}}

\put(0.5,24){\makebox(0,8)[bl]{Example 1: }}
\put(0.5,20){\makebox(0,8)[bl]{Web-based}}
\put(0.5,16){\makebox(0,8)[bl]{system for }}
\put(0.5,12){\makebox(0,8)[bl]{provider}}


\put(18.5,06){\line(1,0){18}} \put(18.5,28){\line(1,0){18}}
\put(18.5,06){\line(0,1){22}} \put(36.5,06){\line(0,1){22}}

\put(19,24){\makebox(0,8)[bl]{Example 2: }}
\put(19,20){\makebox(0,8)[bl]{Web-based}}
\put(19,16){\makebox(0,8)[bl]{system for }}
\put(19,12){\makebox(0,8)[bl]{corporate}}
\put(19,08){\makebox(0,8)[bl]{application}}


\put(37,06){\line(1,0){18}} \put(37,28){\line(1,0){18}}
\put(37,06){\line(0,1){22}} \put(55,06){\line(0,1){22}}

\put(37.5,24){\makebox(0,8)[bl]{Example 3: }}
\put(37.5,20){\makebox(0,8)[bl]{Web-based}}
\put(37.5,16){\makebox(0,8)[bl]{system for }}
\put(37.5,12.5){\makebox(0,8)[bl]{academic}}
\put(37.5,08){\makebox(0,8)[bl]{application}}


\put(55.5,06){\line(1,0){18}} \put(55.5,28){\line(1,0){18}}
\put(55.5,06){\line(0,1){22}} \put(73.5,06){\line(0,1){22}}

\put(56,24){\makebox(0,8)[bl]{Example 4:}}
\put(56,20){\makebox(0,8)[bl]{Web-}}
\put(56,16){\makebox(0,8)[bl]{hosting}}
\put(56,12){\makebox(0,8)[bl]{system  }}
\put(56,08){\makebox(0,8)[bl]{\cite{levshar09} }}


\end{picture}
\end{center}

%


\section{Underlaying Problems/Schemes}


%
%
%
%

\subsection{Multicriteria Ranking}

 Let ~\(H = \{ 1,...,i,...,t \}\)~
 be a set of items which are evaluated upon criteria
 ~\(K = \{ 1,...,j,...,d \}\)~
 and ~\(z_{i,j}\) is an estimate
 (quantitative, ordinal) of item ~\(i\) on criterion ~\(j\).
%
%
 The matrix ~\(\{z_{i,j}\}\)
 is a basis to build a partial order on ~\(H\),
 for example through the following generalized scheme:
 (a) pairwise elements comparison
 to get
 a preference (and/or incomparability, equivalence) binary relation,
 (b) building a partial order on \(H\).
 Here the following partial order
 (partition) as linear ordered subsets of ~\(H\)~
 is searched for:~~
 \( H = \cup_{k=1}^{m} H(k)\),~ \(|H(k_{1}) \cap H(k_{2})| = 0\)~~
 if~~ \(k_{1}\neq k_{2}\),
 ~~\( i_{2}\preceq i_{1}\)~~ \(\forall i_{1} \in H(k_{1})\),
 ~\(\forall i_{2} \in H(k_{2})\), ~ \(k_{1} \leq k_{2}\).

  Set ~\(H(k)\) is called layer ~\(k\), and each item
 ~\(i \in H\)~
 gets priority ~\(r_{i}\)
 that equals the number of the corresponding layer.
 This problem belongs to class of ill-structured problems
 by classification of Simon and Newell \cite{simon58}.
 The list of basic techniques for multicriteria selection
 is the following:
   (1) multi-attribute utility analysis
   \cite{kee76};
   (2) multi-criterion decision making
   \cite{kor84};
   (3)
   Analytic Hierarchy Process (AHP) \cite{saa88};
   (4)
   outranking techniques \cite{roy96}; etc.

\subsection{Knapsack Problems}

 The description of knapsack-like problems
 is presented in
(\cite{keller04},
 \cite{mar90}).
 The basic (simplified) knapsack problem formulation is:

%
 \(\max\sum_{i=1}^{m} c_{i} x_{i}\)

 \(s.t.~\sum_{i=1}^{m} a_{i} x_{i} \leq b\),
 ~\(x_{i} \in \{0,1\}\), ~ \(i=\overline{1,m}\),

%
 where \(x_{i}=1\) if item \(i\) is selected,
 \(c_{i}\) is a value ("utility") for item \(i\), and
 \(a_{i}\) is a weight (or resource required).
 Often nonnegative coefficients are assumed.
 The problem is NP-hard
 and it is
 presented, for example
 in
  (\cite{gar79},
 \cite{mar90}).
 This problem
 can be solved by enumerative methods
 (e.g., Branch-and-Bound, dynamic programming),
 approximate schemes with a limited
 relative error,
 for example, the algorithms are described in
  (\cite{keller04},
  \cite{mar90}).
 In the case of a multiple choice problem,
 the items (e.g., actions) are divided into groups
 and we select elements from each group
 while taking into account a total resource constraint (or constraints):

 \(\max\sum_{i=1}^{m} \sum_{j=1}^{q_{i}} c_{ij} x_{ij}\)

  \(s.t.~\sum_{i=1}^{m} \sum_{j=1}^{q_{i}} a_{ij} x_{ij} \leq
 b,\)

  \(\sum_{j=1}^{q_{i}} x_{ij}=1\), ~\(i=\overline{1,m}\),
 ~~\(x_{ij} \in \{0,1\}\).
%
%


%
 In the case of multicriteria description
 each element (i.e., \((i,j)\)) has a vector profit:

  ~\( \overline{c_{i,j}}    = (  c^{1}_{i,j}, ..., c^{\xi}_{i,j}, ... , c^{r}_{i,j} )
  \).

  A version of multicriteria multiple choice problem
 was
 presented in
  \cite{levs06}):

 \(\max\sum_{i=1}^{m} \sum_{j=1}^{q_{i}} c^{\xi}_{ij} x_{ij}, ~~ \forall \xi = \overline{1,r}\)

 \(s.t.~\sum_{i=1}^{m} \sum_{j=1}^{q_{i}} a_{ij} x_{ij} \leq
 b,\)

 \(\sum_{j=1}^{q_{i}} x_{ij}=1\), \(i=\overline{1,m}\),
 ~~\(x_{ij} \in \{0, 1\}\).

 Evidently, in this case it is reasonable to search for
 Pareto-efficient  solutions
 (by the vector objective function above).
%
%
  Here
 the following solving schemes
 can be used  \cite{levs06}:
 ~(i)
  dynamic programming,
 ~(ii) heuristic based on preliminary multicriteria ranking of elements
 to get their priorities
 and
 step-by-step packing the knapsack (i.e., greedy approach),
 ~(iii) multicriteria ranking of elements to get their ordinal
 priorities and usage of approximate solving scheme (as for knapsack
 problem) based on discrete space of system excellence
 (i.e., lattice as in HMMD).
%
 In the article, greedy heuristic above is used later.

\subsection{Morphological Design}


 Hierarchical Morphological Multicriteria Design (HMMD) approach,
 suggested by Levin
  (e.g., \cite{lev98}, \cite{lev05}, \cite{lev06}),
 is based on the morphological clique problem.
%
%
%
 The composite (modular, decomposable) system
 under examination
 consists
 of the components and their interconnections or compatibilities.
 Basic assumptions of HMMD are the following:
 ~(a) a tree-like structure of the system;
 ~(b) a composite estimate for system quality
     that integrates components (subsystems, parts) qualities and
     qualities of interconnections
      (hereinafter referred as 'IC')
     across subsystems;
 ~(c) monotonic criteria for the system and its components;
 and
 ~(d) quality of system components and IC are evaluated
   on the basis of coordinated ordinal scales.
 The designations are:
  ~(1) design alternatives (DAs) for
  nodes of the model;
  ~(2) priorities of DAs (\(r=\overline{1,k}\);
      \(1\) corresponds to the best level);
  ~(3) ordinal compatibility estimates for each pair of DAs
  (\(w=\overline{0,l}\); \(l\) corresponds to the best level).
 The basic phases of HMMD are (Fig. 3):
  ~{\it 1.} design of the tree-like system model (a preliminary phase);
  ~{\it 2.} generating DAs for model's leaf nodes;
  ~{\it 3.} hierarchical selection and composing of DAs into composite
    DAs for the corresponding higher level of the system
    hierarchy (morphological clique problem); and
  ~{\it 4.} analysis and improvement of the resultant composite DAs (decisions).

\begin{center}
\begin{picture}(70,88)
\put(10,0){\makebox(0,0)[bl] {Fig. 3. 'Bottom-Up' scheme}}

\put(0,11){\line(1,0){26}} \put(0,5){\line(1,0){26}}
\put(0,5){\line(0,1){6}} \put(26,5){\line(0,1){6}}

\put(13,11){\vector(0,1){4}}

\put(0.5,6){\makebox(0,0)[bl]{Generating DAs}}


\put(0,22){\line(1,0){26}} \put(0,15){\line(1,0){26}}
\put(0,15){\line(0,1){7}} \put(26,15){\line(0,1){7}}

\put(13,22){\vector(0,1){4}}

\put(0.5,19){\makebox(0,0)[bl]{Assessment of}}
\put(0.5,16){\makebox(0,0)[bl]{~~~~~~~ DAs}}


\put(25,23.5){\makebox(0,0)[bl]{.~.~.}}

\put(25,34.5){\makebox(0,0)[bl]{.~.~.}}

\put(0,32){\line(1,0){26}} \put(0,26){\line(1,0){26}}
\put(0,26){\line(0,1){6}} \put(26,26){\line(0,1){6}}

\put(13,32){\vector(1,1){5}}

\put(0.5,27){\makebox(0,0)[bl]{Ranking of DAs}}
\put(29,11){\line(1,0){26}} \put(29,5){\line(1,0){26}}
\put(29,5){\line(0,1){6}} \put(55,5){\line(0,1){6}}

\put(42,11){\vector(0,1){4}}

\put(29.5,6){\makebox(0,0)[bl]{Generating DAs}}


\put(29,22){\line(1,0){26}} \put(29,15){\line(1,0){26}}
\put(29,15){\line(0,1){7}} \put(55,15){\line(0,1){7}}

\put(42,22){\vector(0,1){4}}

\put(29.5,19){\makebox(0,0)[bl]{Assessment of}}
\put(29.5,16){\makebox(0,0)[bl]{~~~~~~~ DAs}}

\put(29,32){\line(1,0){26}} \put(29,26){\line(1,0){26}}
\put(29,26){\line(0,1){6}} \put(55,26){\line(0,1){6}}

\put(42,32){\vector(-1,1){5}}

\put(29.5,27){\makebox(0,0)[bl]{Ranking of DAs}}


\put(10,51){\line(1,0){35}} \put(10,37){\line(1,0){35}}
\put(10,37){\line(0,1){14}} \put(45,37){\line(0,1){14}}

\put(27.5,51){\vector(1,1){5}}

\put(11,47){\makebox(0,0)[bl]{Composition of DAs}}
\put(11,43){\makebox(0,0)[bl]{(morphological clique}}
\put(11,39){\makebox(0,0)[bl]{problem)}}


\put(25,74){\line(1,0){45}} \put(25,64){\line(1,0){45}}
\put(25,64){\line(0,1){10}} \put(70,64){\line(0,1){10}}

\put(36,60){\vector(1,1){4}} \put(59,60){\vector(-1,1){4}}


\put(47.5,60){\vector(0,1){4}}

\put(26,70){\makebox(0,0)[bl]{Composition of DAs (mor-}}
\put(26,66){\makebox(0,0)[bl]{phological clique problem}}

\put(45,57){\makebox(0,0)[bl]{.~.~.}}

\put(25,88){\line(1,0){45}} \put(25,78){\line(1,0){45}}
\put(25,78){\line(0,1){10}} \put(70,78){\line(0,1){10}}

\put(47.5,74){\vector(0,1){4}}

\put(26,84){\makebox(0,0)[bl]{Analysis and improvement}}
\put(26,80){\makebox(0,0)[bl]{of resultant composite DAs}}

\end{picture}
\end{center}

 Let ~\(S\) be a system consisting of ~\(m\) parts (components):
 ~\(P(1),...,P(i),...,P(m)\).
 A set of design alternatives
 is generated for each system part above.
  The problem is:

 {\it Find a composite design alternative}
 ~~ \(S=S(1)\star ...\star S(i)\star ...\star S(m)\)~~
 {\it of}~ DAs ({\it one representative design alternative} \(S(i)\)
 {\it for each system component/part} ~\(P(i)\), \(i=\overline{1,m}\))~
 {\it with non-zero}~ IC ~{\it estimates between design alternatives.}

  A discrete space of the system excellence on the basis of the
 following vector is used:
 ~~\(N(S)=(w(S);n(S))\),
 ~where \(w(S)\) is the minimum of pairwise compatibility
 between DAs which correspond to different system components
 (i.e.,
 \(~\forall ~P_{j_{1}}\) and \( P_{j_{2}}\),
 \(1 \leq j_{1} \neq j_{2} \leq m\))
 in \(S\),
 ~\(n(S)=(n_{1},...,n_{r},...n_{k})\),
 ~where ~\(n_{r}\) is the number of DAs of the \(r\)th quality in ~\(S\)
 ~(\(\sum^{k}_{r=1} n_{r} = m\)).
 As a result,
 we search for composite system decisions which are nondominated by ~\(N(S)\)
 (Fig. 4 and Fig. 5).
%
%
 %
 Here an enumerative solving scheme
 (e.g., dynamic programming)
  is used (usually \(m \leq 6\)) \cite{lev98}.

\begin{center}
\begin{picture}(80,80)

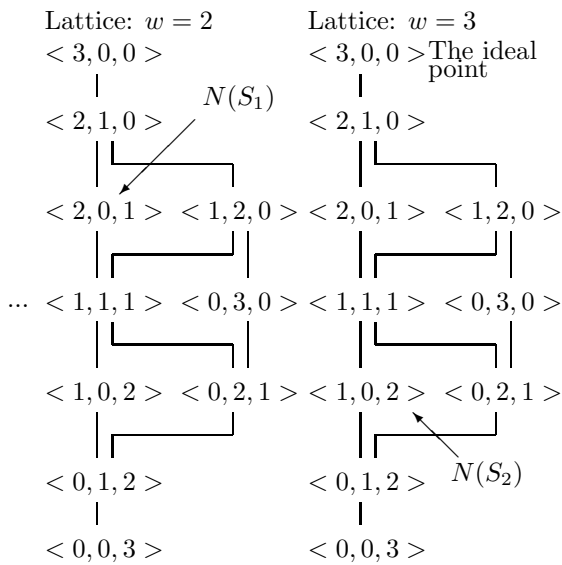
\put(13,0){\makebox(0,0)[bl] {Fig. 4. Lattice of system quality}}


\put(05,76){\makebox(0,0)[bl]{Lattice: \(w=2\) }}

\put(05,71){\makebox(0,0)[bl]{\(<3,0,0>\) }}

\put(12,67){\line(0,1){3}}
\put(05,62){\makebox(0,0)[bl]{\(<2,1,0>\)}}

\put(26,65){\makebox(0,0)[bl]{\(N(S_{1})\)}}
\put(25,64){\vector(-1,-1){10}}

\put(12,55){\line(0,1){6}}
\put(05,50){\makebox(0,0)[bl]{\(<2,0,1>\) }}

\put(12,43){\line(0,1){6}}
\put(05,38){\makebox(0,0)[bl]{\(<1,1,1>\) }}

\put(12,31){\line(0,1){6}}
\put(05,26){\makebox(0,0)[bl]{\(<1,0,2>\) }}


\put(00,39){\makebox(0,0)[bl]{... }}


\put(12,19){\line(0,1){6}}
\put(05,14){\makebox(0,0)[bl]{\(<0,1,2>\) }}

\put(12,10){\line(0,1){3}}

\put(05,05){\makebox(0,0)[bl]{\(<0,0,3>\) }}

\put(14,58){\line(0,1){3}} \put(30,58){\line(-1,0){16}}
\put(30,55){\line(0,1){3}}

\put(23,50){\makebox(0,0)[bl]{\(<1,2,0>\) }}

\put(30,49){\line(0,-1){3}} \put(30,46){\line(-1,0){16}}
\put(14,46){\line(0,-1){3}}
\put(32,43){\line(0,1){6}}
\put(23,38){\makebox(0,0)[bl]{\(<0,3,0>\) }}

\put(14,34){\line(0,1){3}} \put(30,34){\line(-1,0){16}}
\put(30,31){\line(0,1){3}}

\put(32,31){\line(0,1){6}}
\put(23,26){\makebox(0,0)[bl]{\(<0,2,1>\) }}

\put(30,25){\line(0,-1){3}} \put(30,22){\line(-1,0){16}}
\put(14,22){\line(0,-1){3}}


\put(40,76){\makebox(0,0)[bl]{Lattice: \(w=3\) }}

\put(56,72){\makebox(0,0)[bl]{The ideal}}
\put(56,69){\makebox(0,0)[bl]{point}}

\put(40,71){\makebox(0,0)[bl]{\(<3,0,0>\) }}

\put(47,67){\line(0,1){3}}
\put(40,62){\makebox(0,0)[bl]{\(<2,1,0>\)}}

\put(47,55){\line(0,1){6}}
\put(40,50){\makebox(0,0)[bl]{\(<2,0,1>\) }}

\put(47,43){\line(0,1){6}}
\put(40,38){\makebox(0,0)[bl]{\(<1,1,1>\) }}

\put(47,31){\line(0,1){6}}
\put(40,26){\makebox(0,0)[bl]{\(<1,0,2>\) }}


\put(47,19){\line(0,1){6}}
\put(40,14){\makebox(0,0)[bl]{\(<0,1,2>\) }}

\put(47,10){\line(0,1){3}}

\put(40,05){\makebox(0,0)[bl]{\(<0,0,3>\) }}

\put(49,58){\line(0,1){3}} \put(65,58){\line(-1,0){16}}
\put(65,55){\line(0,1){3}}

\put(58,50){\makebox(0,0)[bl]{\(<1,2,0>\) }}

\put(65,49){\line(0,-1){3}} \put(65,46){\line(-1,0){16}}
\put(49,46){\line(0,-1){3}}
\put(67,43){\line(0,1){6}}
\put(58,38){\makebox(0,0)[bl]{\(<0,3,0>\) }}

\put(49,34){\line(0,1){3}} \put(65,34){\line(-1,0){16}}
\put(65,31){\line(0,1){3}}

\put(67,31){\line(0,1){6}}
\put(58,26){\makebox(0,0)[bl]{\(<0,2,1>\) }}

\put(65,25){\line(0,-1){3}} \put(65,22){\line(-1,0){16}}
\put(49,22){\line(0,-1){3}}


\put(59,15){\makebox(0,0)[bl]{\(N(S_{2})\)}}
\put(60,19){\vector(-1,1){6}}

\end{picture}
\end{center}
\begin{center}
\begin{picture}(60,63)
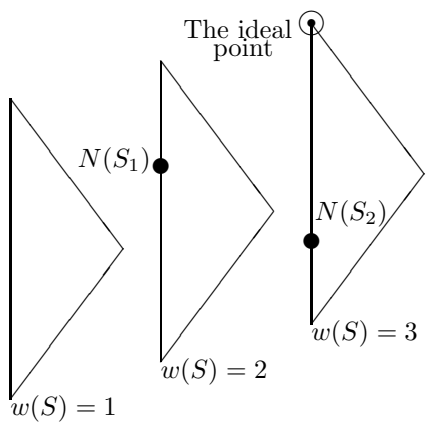
\put(0,0){\makebox(0,0)[bl]{Fig. 5. Illustration of system quality
space}}

\put(0,010){\line(0,1){40}} \put(0,010){\line(3,4){15}}
\put(0,050){\line(3,-4){15}}

\put(20,015){\line(0,1){40}} \put(20,015){\line(3,4){15}}
\put(20,055){\line(3,-4){15}}

\put(40,020){\line(0,1){40}} \put(40,020){\line(3,4){15}}
\put(40,060){\line(3,-4){15}}

\put(40,31){\circle*{2}}
\put(40.5,33){\makebox(0,0)[bl]{\(N(S_{2})\)}}


\put(20,41){\circle*{2}}
\put(9,40){\makebox(0,0)[bl]{\(N(S_{1})\)}}


\put(40,60){\circle*{1}} \put(40,60){\circle{3}}


\put(23,58){\makebox(0,0)[bl]{The ideal}}
\put(27,55){\makebox(0,0)[bl]{point}}


\put(0,7){\makebox(0,0)[bl]{\(w(S)=1\)}}
\put(20,12){\makebox(0,0)[bl]{\(w(S)=2\)}}
\put(40,17){\makebox(0,0)[bl]{\(w(S)=3\)}}
\end{picture}
\end{center}

    Generally, the following layers of system excellence can be considered:
  ~(i) ideal point;
  ~(ii) Pareto-efficient points;
  ~(iii) a neighborhood of Pareto-efficient DAs
 (e.g., a composite decision of this set can be
 transformed into a Pareto-efficient point on the basis of an
 improvement action(s)).
 Clearly, the compatibility component of vector ~\(N(S)\)
 can be considered on the basis of a poset-like scale too
 (as \(n(S)\))
  (\cite{levf01}, \cite{lev06}).
 In this case, the discrete space of
 system excellence will be an analogical lattice.
%

%
%
 Fig. 6 and Fig. 7 illustrate the composition problem
 (by a numerical example for a system consisting of three parts
 ~\(S = X \star Y \star Z\)).
 Priorities of DAs are shown in Fig. 6 in parentheses and are
 depicted in Fig. 7;
  compatibility estimates are pointed out in Fig. 7).
 In the example, the resultant composite decisions are
 (Fig. 4, Fig. 5, Fig. 6, Fig. 7):
 ~\(S_{1}=X_{2}\star Y_{1}\star Z_{2}\), ~\(N(S_{1}) = (2;2,0,1)\);
 ~\(S_{2}=X_{3}\star Y_{1}\star Z_{3}\), ~\(N(S_{2}) = (3;1,0,2)\).

\begin{center}
\begin{picture}(46,39)

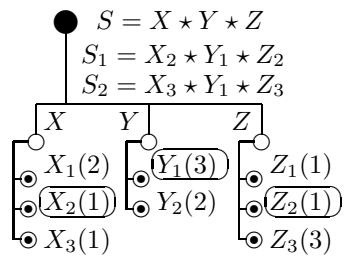
\put(0,0){\makebox(0,0)[bl] {Fig. 6. Example of composition}}

\put(4,5){\makebox(0,8)[bl]{\(X_{3}(1)\)}}
\put(4,10){\makebox(0,8)[bl]{\(X_{2}(1)\)}}
\put(4,15){\makebox(0,8)[bl]{\(X_{1}(2)\)}}

\put(8.5,12){\oval(10,4)}


\put(19,10){\makebox(0,8)[bl]{\(Y_{2}(2)\)}}
\put(19,15){\makebox(0,8)[bl]{\(Y_{1}(3)\)}}

\put(23.5,17){\oval(10,4)}

\put(34,5){\makebox(0,8)[bl]{\(Z_{3}(3)\)}}
\put(34,10){\makebox(0,8)[bl]{\(Z_{2}(1)\)}}
\put(34,15){\makebox(0,8)[bl]{\(Z_{1}(1)\)}}

\put(38.5,12){\oval(10,4)}

\put(3,20){\circle{2}} \put(18,20){\circle{2}}
\put(33,20){\circle{2}}

\put(0,20){\line(1,0){02}} \put(15,20){\line(1,0){02}}
\put(30,20){\line(1,0){02}}

\put(0,20){\line(0,-1){13}} \put(15,20){\line(0,-1){09}}
\put(30,20){\line(0,-1){13}}

\put(30,15){\line(1,0){01}} \put(30,11){\line(1,0){01}}
\put(30,7){\line(1,0){01}}

\put(32,15){\circle{2}} \put(32,15){\circle*{1}}
\put(32,11){\circle{2}} \put(32,11){\circle*{1}}
\put(32,7){\circle{2}} \put(32,7){\circle*{1}}


\put(15,11){\line(1,0){01}} \put(15,15){\line(1,0){01}}

\put(17,11){\circle{2}} \put(17,11){\circle*{1}}


\put(17,15){\circle{2}}


\put(17,15){\circle*{1}}


\put(0,7){\line(1,0){01}} \put(0,11){\line(1,0){01}}
\put(0,15){\line(1,0){01}}

\put(2,11){\circle{2}} \put(2,15){\circle{2}}
\put(2,11){\circle*{1}} \put(2,15){\circle*{1}}
\put(2,7){\circle{2}} \put(2,7){\circle*{1}}

\put(3,25){\line(0,-1){04}} \put(18,25){\line(0,-1){04}}
\put(33,25){\line(0,-1){04}}

\put(3,25){\line(1,0){30}}

\put(7,25){\line(0,1){10}} \put(7,36){\circle*{3}}

\put(04,21.5){\makebox(0,8)[bl]{\(X\) }}
\put(14,21.5){\makebox(0,8)[bl]{\(Y\) }}
\put(29,21.5){\makebox(0,8)[bl]{\(Z\) }}

\put(11,35){\makebox(0,8)[bl]{\(S = X \star Y \star Z \) }}

\put(9,30){\makebox(0,8)[bl]{\(S_{1}=X_{2}\star Y_{1}\star
Z_{2}\)}}

\put(9,26){\makebox(0,8)[bl]{\(S_{2}=X_{3}\star Y_{1}\star
Z_{3}\)}}

\end{picture}
\end{center}

\begin{center}
\begin{picture}(50,38)
\put(0,0){\makebox(0,0)[bl] {Fig. 7. Concentric presentation}}

\put(1,10){\line(0,1){6}} \put(7,10){\line(0,1){6}}

\put(01,10){\line(1,0){18}} \put(01,16){\line(1,0){18}}

\put(2,12){\makebox(0,0)[bl]{\(Z_{3}\)}} \put(8,10){\line(0,1){6}}
\put(9,12){\makebox(0,0)[bl]{\(Z_{2}\)}}

\put(11,13){\oval(5,8)}

\put(14,12){\makebox(0,0)[bl]{\(Z_{1}\)}}
\put(19,10){\line(0,1){6}}

\put(25,10){\line(1,0){14}} \put(25,16){\line(1,0){14}}
\put(25,10){\line(0,1){6}}
\put(27,10){\line(0,1){6}}
\put(28,12){\makebox(0,0)[bl]{\(Y_{2}\)}}
\put(33,10){\line(0,1){6}}
\put(34,12){\makebox(0,0)[bl]{\(Y_{1}\)}}
\put(39,10){\line(0,1){6}}

\put(36,13){\oval(5,8)}

\put(19,20){\line(0,1){17}} \put(25,20){\line(0,1){17}}
\put(19,20){\line(1,0){6}}
\put(20,21.5){\makebox(0,0)[bl]{\(X_{3}\)}}
\put(20,26){\makebox(0,0)[bl]{\(X_{2}\)}}

\put(22,27.5){\oval(8,4)}

 \put(19,30){\line(1,0){6}}
\put(20,32){\makebox(0,0)[bl]{\(X_{1}\)}}
\put(19,36){\line(1,0){6}} \put(19,37){\line(1,0){6}}

\put(10,10){\line(0,-1){3}} \put(10,7){\line(1,0){26}}
\put(36,7){\line(0,1){3}}

\put(37,6){\makebox(0,0)[bl]{\(2\)}}


\put(10,16){\line(0,1){11}} \put(10,27){\line(1,0){09}}
\put(11,29){\makebox(0,0)[bl]{\(2\)}}

\put(36,16){\line(0,1){11}} \put(36,27){\line(-1,0){11}}
\put(37,22){\makebox(0,0)[bl]{\(3\)}}



\put(4,16){\line(0,1){7}} \put(4,23){\line(1,0){15}}
\put(4,24){\makebox(0,0)[bl]{\(3\)}}


\put(04,10){\line(0,-1){5}} \put(04,5){\line(1,0){31}}
\put(35,5){\line(0,1){5}}

\put(01,6){\makebox(0,0)[bl]{\(3\)}}


\put(35,16){\line(0,1){07}} \put(35,23){\line(-1,0){10}}
\put(32,19){\makebox(0,0)[bl]{\(3\)}}

\end{picture}
\end{center}

\section{Applied Web-based System}

 The structure
 (infrastructure)
 of
 an applied Web based system
 is examined
 as a combination of
 two main parts:
 software and hardware.
 The basic example is targeted to a communication service provider
 (example 1).

\subsection{Hierarchical Model and Components}

 The tree-like model of the considered information system infrastructure
 is depicted in Fig. 8.

 DAs for system components are the following:

 (1) server for DBs \(M\):
    PC (\(M_{1}\)),
    Supermicro (\(M_{2}\)), and
    Sun (\(M_{3}\));

 (2) server for applications \(E\):
    on server of DBs (\(E_{1}\)),
    Sun (\(E_{2}\)),
    Supermicro (\(E_{3}\)), and
    PC (\(E_{4}\));

 (3) Web-server \(W\):
    Apache HTTP-server (\(W_{1}\)),
    Microsoft IIS (\(W_{2}\)),
    Bea Weblogic (\(W_{3}\)),
    Web Sphere (\(W_{4}\)), and
    Weblogic cluster (\(W_{5}\));

 (4) DBMS \(D\):
    Oracle (\(D_{1}\)),
    Microsoft SQL (\(D_{2}\)), and
    designed SQL (\(D_{3}\)); and

 (5) operation system \(O\):
    Windows 2000 server (\(O_{1}\)),
    Windows 2003 (\(O_{2}\)),
    Solaris (\(O_{3}\)),
    FreeBSD (\(O_{4}\)), and
    RHEL AS (\(O_{5}\)).

\begin{center}
\begin{picture}(70,53)

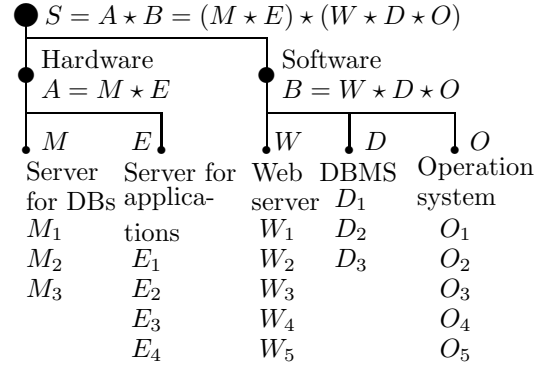
\put(0,0){\makebox(0,0)[bl] {Fig. 8. Structure
 of
 applied
 Web-based
 system}}

\put(00,51){\circle*{3}}

\put(2.5,49){\makebox(0,0)[bl]{\(S=A\star B = ( M\star E)\star
(W\star D\star O) \)}}

\put(02,44){\makebox(0,0)[bl]{Hardware}}
\put(02,40){\makebox(0,0)[bl]{\(A=M\star E\)}}

\put(34,44){\makebox(0,0)[bl]{Software }}
\put(34,40){\makebox(0,0)[bl]{\(B=W\star D\star O\)}}

\put(00,46){\line(0,1){4}} \put(00,48){\line(1,0){32}}
\put(32,43){\line(0,1){05}}

\put(32,43){\circle*{2}}

\put(32,33){\line(0,1){10}}

\put(00,43){\line(0,1){05}}

\put(00,43){\circle*{2}} \put(00,42){\line(0,1){6}}

\put(00,38){\line(1,0){18}} \put(32,38){\line(1,0){25}}

\put(00,33){\line(0,1){10}} \put(18,33){\line(0,1){05}}
\put(32,33){\line(0,1){05}}

\put(43,33){\line(0,1){05}} \put(57,33){\line(0,1){05}}

\put(00,33){\circle*{1}} \put(18,33){\circle*{1}}

\put(32,33){\circle*{1}} \put(43,33){\circle*{1}}
\put(57,33){\circle*{1}}

\put(02,33){\makebox(0,0)[bl]{\(M\)}}
\put(14,33){\makebox(0,0)[bl]{\(E\)}}
\put(33,33){\makebox(0,0)[bl]{\(W\)}}
\put(45,33){\makebox(0,0)[bl]{\(D\)}}
\put(59,33){\makebox(0,0)[bl]{\(O\)}}

\put(52,29){\makebox(0,0)[bl]{Operation}}
\put(52,25){\makebox(0,0)[bl]{system}}
\put(55,21){\makebox(0,0)[bl]{\(O_{1}\)}}
\put(55,17){\makebox(0,0)[bl]{\(O_{2}\)}}
\put(55,13){\makebox(0,0)[bl]{\(O_{3}\)}}
\put(55,9){\makebox(0,0)[bl]{\(O_{4}\)}}
\put(55,5){\makebox(0,0)[bl]{\(O_{5}\)}}

\put(39,29){\makebox(0,0)[bl]{DBMS}}
\put(41,25){\makebox(0,0)[bl]{\(D_{1}\)}}
\put(41,21){\makebox(0,0)[bl]{\(D_{2}\)}}
\put(41,17){\makebox(0,0)[bl]{\(D_{3}\)}}

\put(30,29){\makebox(0,0)[bl]{Web}}
\put(30,25){\makebox(0,0)[bl]{server}}
\put(31,21){\makebox(0,0)[bl]{\(W_{1}\)}}
\put(31,17){\makebox(0,0)[bl]{\(W_{2}\)}}
\put(31,13){\makebox(0,0)[bl]{\(W_{3}\)}}
\put(31,9){\makebox(0,0)[bl]{\(W_{4}\)}}
\put(31,5){\makebox(0,0)[bl]{\(W_{5}\)}}

\put(13,29){\makebox(0,0)[bl]{Server for}}
\put(13,25){\makebox(0,0)[bl]{applica-}}
\put(13,21){\makebox(0,0)[bl]{tions}}
\put(14,17){\makebox(0,0)[bl]{\(E_{1}\)}}
\put(14,13){\makebox(0,0)[bl]{\(E_{2}\)}}
\put(14,9){\makebox(0,0)[bl]{\(E_{3}\)}}
\put(14,05){\makebox(0,0)[bl]{\(E_{4}\)}}

\put(00,29){\makebox(0,0)[bl]{Server }}
\put(00,25){\makebox(0,0)[bl]{for DBs}}
\put(00,21){\makebox(0,0)[bl]{\(M_{1}\)}}
\put(00,17){\makebox(0,0)[bl]{\(M_{2}\)}}
\put(00,13){\makebox(0,0)[bl]{\(M_{3}\)}}

\end{picture}
\end{center}

 \subsection{Assessment}

 The following criteria are used for assessment of DAs
 ('+' corresponds to positive orientation of an ordinal scale as
 \([1,6]\)
 when the biggest estimate is the best
 one,
  '-' corresponds to the negative orientation of the scale
 when the smallest estimate is the best
 one):
 {\it (a)} cost \(C_{1}\) ('-'),
 {\it (b)} performance \(C_{2}\) ('+'),
 {\it (c)} complexity of maintenance \(C_{3}\) ('-'), and
 {\it (d)} scalability \(C_{4}\) ('+').
 The corresponding estimates for DA \(i\)
 are as follows ~\(z_{i} = (z_{i1},z_{i2},z_{i3},z_{i4})\).

 Tables 2 and 3 contain ordinal estimates of DAs upon the
 above-mentioned criteria (expert judgment).
 Criteria weights for three application examples are contained in Table 4.
%
%
%
%
  Estimates of compatibility between DAs are contained
  in Tables 5 and 6 (expert judgment).

\subsection{Communication Service Provider}

 The resultant priorities of DAs are obtained as result of
 multicriteria ranking (Electre-like method).
 The priorities of DAs for example 1 (communication service provider)
  are shown in Fig. 9 in parentheses.

\begin{center}
\begin{picture}(30,68)
\put(00,64){\makebox(0,0)[bl]{Table 2. Estimates}}

\put(00,0){\line(1,0){26}} \put(00,50){\line(1,0){26}}
\put(10,56){\line(1,0){16}} \put(00,62){\line(1,0){26}}

\put(00,0){\line(0,1){62}} \put(10,00){\line(0,1){62}}
\put(26,0){\line(0,1){62}}

\put(14,50){\line(0,1){6}} \put(18,50){\line(0,1){6}}
\put(22,50){\line(0,1){6}}

\put(01,58){\makebox(0,0)[bl]{DAs}}
\put(12,58){\makebox(0,0)[bl]{Criteria}}

\put(11,52){\makebox(0,0)[bl]{\(1\)}}
\put(15,52){\makebox(0,0)[bl]{\(2\)}}
\put(19,52){\makebox(0,0)[bl]{\(3\)}}
\put(23,52){\makebox(0,0)[bl]{\(4\)}}


\put(01,46){\makebox(0,0)[bl]{\(M_{1}\)}}
\put(01,42){\makebox(0,0)[bl]{\(M_{2}\)}}
\put(01,38){\makebox(0,0)[bl]{\(M_{3}\)}}
\put(01,34){\makebox(0,0)[bl]{\(E_{1}\)}}
\put(01,30){\makebox(0,0)[bl]{\(E_{2}\)}}
\put(01,26){\makebox(0,0)[bl]{\(E_{3}\)}}
\put(01,22){\makebox(0,0)[bl]{\(E_{4}\)}}
\put(01,18){\makebox(0,0)[bl]{\(W_{1}\)}}
\put(01,14){\makebox(0,0)[bl]{\(W_{2}\)}}
\put(01,10){\makebox(0,0)[bl]{\(W_{3}\)}}
\put(01,06){\makebox(0,0)[bl]{\(W_{4}\)}}
\put(01,02){\makebox(0,0)[bl]{\(W_{5}\)}}

\put(11,46){\makebox(0,0)[bl]{\(2\)}}
\put(15,46){\makebox(0,0)[bl]{\(2\)}}
\put(19,46){\makebox(0,0)[bl]{\(3\)}}
\put(23,46){\makebox(0,0)[bl]{\(2\)}}

\put(11,42){\makebox(0,0)[bl]{\(5\)}}
\put(15,42){\makebox(0,0)[bl]{\(4\)}}
\put(19,42){\makebox(0,0)[bl]{\(4\)}}
\put(23,42){\makebox(0,0)[bl]{\(3\)}}

\put(11,38){\makebox(0,0)[bl]{\(6\)}}
\put(15,38){\makebox(0,0)[bl]{\(5\)}}
\put(19,38){\makebox(0,0)[bl]{\(5\)}}
\put(23,38){\makebox(0,0)[bl]{\(5\)}}

\put(11,34){\makebox(0,0)[bl]{\(1\)}}
\put(15,34){\makebox(0,0)[bl]{\(2\)}}
\put(19,34){\makebox(0,0)[bl]{\(3\)}}
\put(23,34){\makebox(0,0)[bl]{\(1\)}}

\put(11,30){\makebox(0,0)[bl]{\(6\)}}
\put(15,30){\makebox(0,0)[bl]{\(5\)}}
\put(19,30){\makebox(0,0)[bl]{\(5\)}}
\put(23,30){\makebox(0,0)[bl]{\(5\)}}

\put(11,26){\makebox(0,0)[bl]{\(5\)}}
\put(15,26){\makebox(0,0)[bl]{\(4\)}}
\put(19,26){\makebox(0,0)[bl]{\(4\)}}
\put(23,26){\makebox(0,0)[bl]{\(3\)}}

\put(11,22){\makebox(0,0)[bl]{\(2\)}}
\put(15,22){\makebox(0,0)[bl]{\(2\)}}
\put(19,22){\makebox(0,0)[bl]{\(3\)}}
\put(23,22){\makebox(0,0)[bl]{\(2\)}}

\put(11,18){\makebox(0,0)[bl]{\(1\)}}
\put(15,18){\makebox(0,0)[bl]{\(5\)}}
\put(19,18){\makebox(0,0)[bl]{\(2\)}}
\put(23,18){\makebox(0,0)[bl]{\(2\)}}

\put(11,14){\makebox(0,0)[bl]{\(4\)}}
\put(15,14){\makebox(0,0)[bl]{\(3\)}}
\put(19,14){\makebox(0,0)[bl]{\(3\)}}
\put(23,14){\makebox(0,0)[bl]{\(5\)}}

\put(11,10){\makebox(0,0)[bl]{\(5\)}}
\put(15,10){\makebox(0,0)[bl]{\(3\)}}
\put(19,10){\makebox(0,0)[bl]{\(4\)}}
\put(23,10){\makebox(0,0)[bl]{\(4\)}}

\put(11,06){\makebox(0,0)[bl]{\(4\)}}
\put(15,06){\makebox(0,0)[bl]{\(4\)}}
\put(19,06){\makebox(0,0)[bl]{\(5\)}}
\put(23,06){\makebox(0,0)[bl]{\(3\)}}

\put(11,02){\makebox(0,0)[bl]{\(6\)}}
\put(15,02){\makebox(0,0)[bl]{\(3\)}}
\put(19,02){\makebox(0,0)[bl]{\(4\)}}
\put(23,02){\makebox(0,0)[bl]{\(5\)}}

\end{picture}
%
\begin{picture}(30,52)
\put(00,48){\makebox(0,0)[bl]{Table 3. Estimates}}

\put(00,0){\line(1,0){26}} \put(00,34){\line(1,0){26}}
\put(10,40){\line(1,0){16}} \put(00,46){\line(1,0){26}}

\put(00,0){\line(0,1){46}} \put(10,00){\line(0,1){46}}
\put(26,0){\line(0,1){46}}

\put(14,34){\line(0,1){6}} \put(18,34){\line(0,1){6}}
\put(22,34){\line(0,1){6}}

\put(01,42){\makebox(0,0)[bl]{DAs}}
\put(12,42){\makebox(0,0)[bl]{Criteria}}

\put(11,36){\makebox(0,0)[bl]{\(1\)}}
\put(15,36){\makebox(0,0)[bl]{\(2\)}}
\put(19,36){\makebox(0,0)[bl]{\(3\)}}
\put(23,36){\makebox(0,0)[bl]{\(4\)}}


\put(01,30){\makebox(0,0)[bl]{\(D_{1}\)}}
\put(01,26){\makebox(0,0)[bl]{\(D_{2}\)}}
\put(01,22){\makebox(0,0)[bl]{\(D_{3}\)}}
\put(01,18){\makebox(0,0)[bl]{\(O_{1}\)}}
\put(01,14){\makebox(0,0)[bl]{\(O_{2}\)}}
\put(01,10){\makebox(0,0)[bl]{\(O_{3}\)}}
\put(01,06){\makebox(0,0)[bl]{\(O_{4}\)}}
\put(01,02){\makebox(0,0)[bl]{\(O_{5}\)}}

\put(11,30){\makebox(0,0)[bl]{\(6\)}}
\put(15,30){\makebox(0,0)[bl]{\(4\)}}
\put(19,30){\makebox(0,0)[bl]{\(4\)}}
\put(23,30){\makebox(0,0)[bl]{\(5\)}}

\put(11,26){\makebox(0,0)[bl]{\(5\)}}
\put(15,26){\makebox(0,0)[bl]{\(4\)}}
\put(19,26){\makebox(0,0)[bl]{\(3\)}}
\put(23,26){\makebox(0,0)[bl]{\(4\)}}

\put(11,22){\makebox(0,0)[bl]{\(1\)}}
\put(15,22){\makebox(0,0)[bl]{\(3\)}}
\put(19,22){\makebox(0,0)[bl]{\(3\)}}
\put(23,22){\makebox(0,0)[bl]{\(3\)}}

\put(11,18){\makebox(0,0)[bl]{\(3\)}}
\put(15,18){\makebox(0,0)[bl]{\(2\)}}
\put(19,18){\makebox(0,0)[bl]{\(2\)}}
\put(23,18){\makebox(0,0)[bl]{\(2\)}}

\put(11,14){\makebox(0,0)[bl]{\(4\)}}
\put(15,14){\makebox(0,0)[bl]{\(3\)}}
\put(19,14){\makebox(0,0)[bl]{\(1\)}}
\put(23,14){\makebox(0,0)[bl]{\(4\)}}

\put(11,10){\makebox(0,0)[bl]{\(1\)}}
\put(15,10){\makebox(0,0)[bl]{\(5\)}}
\put(19,10){\makebox(0,0)[bl]{\(5\)}}
\put(23,10){\makebox(0,0)[bl]{\(5\)}}

\put(11,06){\makebox(0,0)[bl]{\(1\)}}
\put(15,06){\makebox(0,0)[bl]{\(4\)}}
\put(19,06){\makebox(0,0)[bl]{\(4\)}}
\put(23,06){\makebox(0,0)[bl]{\(3\)}}

\put(11,02){\makebox(0,0)[bl]{\(1\)}}
\put(15,02){\makebox(0,0)[bl]{\(4\)}}
\put(19,02){\makebox(0,0)[bl]{\(3\)}}
\put(23,02){\makebox(0,0)[bl]{\(4\)}}

\end{picture}
\end{center}

\begin{center}
\begin{picture}(58,32)
\put(07,28){\makebox(0,0)[bl]{Table 4. Criteria weights}}

\put(00,0){\line(1,0){56}} \put(00,14){\line(1,0){56}}
\put(32,20){\line(1,0){24}} \put(00,26){\line(1,0){56}}

\put(00,0){\line(0,1){26}} \put(32,00){\line(0,1){26}}
\put(56,0){\line(0,1){26}}

\put(38,14){\line(0,1){6}} \put(44,14){\line(0,1){6}}
\put(50,14){\line(0,1){6}}

\put(01,22){\makebox(0,0)[bl]{Application}}
\put(01,18){\makebox(0,0)[bl]{example }}

\put(36,22){\makebox(0,0)[bl]{Criteria}}

\put(34,16){\makebox(0,0)[bl]{\(1\)}}
\put(40,16){\makebox(0,0)[bl]{\(2\)}}
\put(46,16){\makebox(0,0)[bl]{\(3\)}}
\put(52,16){\makebox(0,0)[bl]{\(4\)}}

\put(01,10){\makebox(0,0)[bl]{1.Provider}}
\put(01,06){\makebox(0,0)[bl]{2.Corporate system}}
\put(01,02){\makebox(0,0)[bl]{3.Academic system}}

\put(33,10){\makebox(0,0)[bl]{\(-1\)}}
\put(40,10){\makebox(0,0)[bl]{\(1\)}}
\put(45,10){\makebox(0,0)[bl]{\(-1\)}}
\put(52,10){\makebox(0,0)[bl]{\(1\)}}

\put(33,06){\makebox(0,0)[bl]{\(-3\)}}
\put(40,06){\makebox(0,0)[bl]{\(1\)}}
\put(45,06){\makebox(0,0)[bl]{\(-2\)}}
\put(52,06){\makebox(0,0)[bl]{\(1\)}}

\put(33,02){\makebox(0,0)[bl]{\(-1\)}}
\put(40,02){\makebox(0,0)[bl]{\(3\)}}
\put(45,02){\makebox(0,0)[bl]{\(-1\)}}
\put(52,02){\makebox(0,0)[bl]{\(1\)}}

\end{picture}
\end{center}

\begin{center}
\begin{picture}(30,26)
\put(0,22){\makebox(0,0)[bl]{Table 5. Compatibility}}

\put(00,0){\line(1,0){27}} \put(00,14){\line(1,0){27}}
\put(00,20){\line(1,0){27}}

\put(00,0){\line(0,1){20}} \put(07,0){\line(0,1){20}}
\put(27,0){\line(0,1){20}}

\put(01,10){\makebox(0,0)[bl]{\(M_{1}\)}}
\put(01,06){\makebox(0,0)[bl]{\(M_{2}\)}}
\put(01,02){\makebox(0,0)[bl]{\(M_{3}\)}}

\put(12,14){\line(0,1){6}} \put(17,14){\line(0,1){6}}
\put(22,14){\line(0,1){6}}

\put(07.4,16){\makebox(0,0)[bl]{\(E_{1}\)}}
\put(12.4,16){\makebox(0,0)[bl]{\(E_{2}\)}}
\put(17.4,16){\makebox(0,0)[bl]{\(E_{3}\)}}
\put(22.4,16){\makebox(0,0)[bl]{\(E_{4}\)}}

\put(09,10){\makebox(0,0)[bl]{\(3\)}}
\put(14,10){\makebox(0,0)[bl]{\(3\)}}
\put(19,10){\makebox(0,0)[bl]{\(3\)}}
\put(24,10){\makebox(0,0)[bl]{\(3\)}}

\put(09,6){\makebox(0,0)[bl]{\(3\)}}
\put(14,6){\makebox(0,0)[bl]{\(3\)}}
\put(19,6){\makebox(0,0)[bl]{\(3\)}}
\put(24,6){\makebox(0,0)[bl]{\(3\)}}

\put(09,2){\makebox(0,0)[bl]{\(3\)}}
\put(14,2){\makebox(0,0)[bl]{\(3\)}}
\put(19,2){\makebox(0,0)[bl]{\(3\)}}
\put(24,2){\makebox(0,0)[bl]{\(3\)}}

\end{picture}
\end{center}
\begin{center}
\begin{picture}(52,46)
\put(06,42){\makebox(0,0)[bl]{Table 6. Compatibility}}

\put(00,0){\line(1,0){47}} \put(00,34){\line(1,0){47}}
\put(00,40){\line(1,0){47}}

\put(00,0){\line(0,1){40}} \put(07,0){\line(0,1){40}}
\put(47,0){\line(0,1){40}}

\put(01,30){\makebox(0,0)[bl]{\(W_{1}\)}}
\put(01,26){\makebox(0,0)[bl]{\(W_{2}\)}}
\put(01,22){\makebox(0,0)[bl]{\(W_{3}\)}}
\put(01,18){\makebox(0,0)[bl]{\(W_{4}\)}}
\put(01,14){\makebox(0,0)[bl]{\(W_{5}\)}}
\put(01,10){\makebox(0,0)[bl]{\(D_{1}\)}}
\put(01,06){\makebox(0,0)[bl]{\(D_{2}\)}}
\put(01,02){\makebox(0,0)[bl]{\(D_{3}\)}}

\put(12,34){\line(0,1){6}} \put(17,34){\line(0,1){6}}
\put(22,34){\line(0,1){6}} \put(27,34){\line(0,1){6}}
\put(32,34){\line(0,1){6}} \put(37,34){\line(0,1){6}}
\put(42,34){\line(0,1){6}}

\put(07.4,36){\makebox(0,0)[bl]{\(D_{1}\)}}
\put(12.4,36){\makebox(0,0)[bl]{\(D_{2}\)}}
\put(17.4,36){\makebox(0,0)[bl]{\(D_{3}\)}}
\put(22.4,36){\makebox(0,0)[bl]{\(O_{1}\)}}
\put(27.4,36){\makebox(0,0)[bl]{\(O_{2}\)}}
\put(32.4,36){\makebox(0,0)[bl]{\(O_{3}\)}}
\put(37.4,36){\makebox(0,0)[bl]{\(O_{4}\)}}
\put(42.4,36){\makebox(0,0)[bl]{\(O_{5}\)}}

\put(09,30){\makebox(0,0)[bl]{\(3\)}}
\put(14,30){\makebox(0,0)[bl]{\(3\)}}
\put(19,30){\makebox(0,0)[bl]{\(3\)}}

\put(24,30){\makebox(0,0)[bl]{\(2\)}}
\put(29,30){\makebox(0,0)[bl]{\(2\)}}
\put(34,30){\makebox(0,0)[bl]{\(3\)}}
\put(39,30){\makebox(0,0)[bl]{\(3\)}}
\put(44,30){\makebox(0,0)[bl]{\(3\)}}

\put(09,26){\makebox(0,0)[bl]{\(3\)}}
\put(14,26){\makebox(0,0)[bl]{\(3\)}}
\put(19,26){\makebox(0,0)[bl]{\(3\)}}

\put(24,26){\makebox(0,0)[bl]{\(3\)}}
\put(29,26){\makebox(0,0)[bl]{\(3\)}}
\put(34,26){\makebox(0,0)[bl]{\(0\)}}
\put(39,26){\makebox(0,0)[bl]{\(0\)}}
\put(44,26){\makebox(0,0)[bl]{\(0\)}}

\put(09,22){\makebox(0,0)[bl]{\(3\)}}
\put(14,22){\makebox(0,0)[bl]{\(3\)}}
\put(19,22){\makebox(0,0)[bl]{\(3\)}}

\put(24,22){\makebox(0,0)[bl]{\(3\)}}
\put(29,22){\makebox(0,0)[bl]{\(3\)}}
\put(34,22){\makebox(0,0)[bl]{\(3\)}}
\put(39,22){\makebox(0,0)[bl]{\(3\)}}
\put(44,22){\makebox(0,0)[bl]{\(3\)}}

\put(09,18){\makebox(0,0)[bl]{\(3\)}}
\put(14,18){\makebox(0,0)[bl]{\(3\)}}
\put(19,18){\makebox(0,0)[bl]{\(3\)}}

\put(24,18){\makebox(0,0)[bl]{\(0\)}}
\put(29,18){\makebox(0,0)[bl]{\(3\)}}
\put(34,18){\makebox(0,0)[bl]{\(3\)}}
\put(39,18){\makebox(0,0)[bl]{\(0\)}}
\put(44,18){\makebox(0,0)[bl]{\(0\)}}

\put(09,14){\makebox(0,0)[bl]{\(3\)}}
\put(14,14){\makebox(0,0)[bl]{\(3\)}}
\put(19,14){\makebox(0,0)[bl]{\(3\)}}

\put(24,14){\makebox(0,0)[bl]{\(3\)}}
\put(29,14){\makebox(0,0)[bl]{\(3\)}}
\put(34,14){\makebox(0,0)[bl]{\(3\)}}
\put(39,14){\makebox(0,0)[bl]{\(3\)}}
\put(44,14){\makebox(0,0)[bl]{\(3\)}}

\put(24,10){\makebox(0,0)[bl]{\(3\)}}
\put(29,10){\makebox(0,0)[bl]{\(3\)}}
\put(34,10){\makebox(0,0)[bl]{\(3\)}}
\put(39,10){\makebox(0,0)[bl]{\(0\)}}
\put(44,10){\makebox(0,0)[bl]{\(1\)}}

\put(24,6){\makebox(0,0)[bl]{\(3\)}}
\put(29,6){\makebox(0,0)[bl]{\(3\)}}
\put(34,6){\makebox(0,0)[bl]{\(0\)}}
\put(39,6){\makebox(0,0)[bl]{\(0\)}}
\put(44,6){\makebox(0,0)[bl]{\(1\)}}

\put(24,2){\makebox(0,0)[bl]{\(3\)}}
\put(29,2){\makebox(0,0)[bl]{\(3\)}}
\put(34,2){\makebox(0,0)[bl]{\(3\)}}
\put(39,2){\makebox(0,0)[bl]{\(3\)}}
\put(44,2){\makebox(0,0)[bl]{\(3\)}}

\end{picture}
\end{center}

 For system part \(A\),
 we get the following Pareto-efficient composite DA
 (superscript for \(A\), \(B\), and \(S\)
 corresponds to the number of applied application as 1, 2, 3):
 ~\(A^{1}_{1} = M_{2} \star E_{2} \), \(N(A^{1}_{1})= (3;1,1,0)\).
 Fig. 10 illustrates the space of quality for \(N(A^{1}_{1})\).
%
%
%
%
 For system part \(B\),
  we get the following Pareto-efficient composite DAs:
 ~\(B^{1}_{1} = W_{1} \star D_{3} \star O_{3}\), \(N(B^{1}_{1})=
 (3;2,1,0)\);
 ~\(B^{1}_{2} = W_{2} \star D_{2} \star O_{2}\), \(N(B^{1}_{2})=
 (3;2,1,0)\); and
 ~\(B^{1}_{3} = W_{1} \star D_{2} \star O_{5}\), \(N(B^{1}_{3})=
 (1;3,0,0)\).

\begin{center}
\begin{picture}(75,80)
\put(07,0){\makebox(0,0)[bl] {Fig. 9. Communication provider}}


\put(00,78){\circle*{3}}

\put(03,77){\makebox(0,0)[bl]{\(S=A\star B\)}}

\put(02,72){\makebox(0,0)[bl] {\(S^{1}_{1}=A^{1}_{1}\star
B^{1}_{1}=(M_{2}\star E_{2})\star (W_{1}\star D_{3}\star
O_{3})\)}}

\put(02,68){\makebox(0,0)[bl] {\(S^{1}_{2}=A^{1}_{1}\star
B^{1}_{2}=(M_{2}\star E_{2})\star (W_{2}\star D_{2}\star
O_{2})\)}}

\put(02,64){\makebox(0,0)[bl] {\(S^{1}_{3}=A^{1}_{1}\star
B^{1}_{3}=(M_{2}\star E_{2})\star (W_{1}\star D_{2}\star
O_{5})\)}}

\put(02,60){\makebox(0,0)[bl]{Hardware }}
\put(02,56){\makebox(0,0)[bl]{\(A=M\star E\)}}

\put(33,60){\makebox(0,0)[bl]{Software}}
\put(33,56){\makebox(0,0)[bl]{\(B=W\star D\star O\)}}

\put(00,54){\line(0,1){23}} \put(00,54){\line(1,0){32}}
\put(32,49){\line(0,1){05}} \put(32,49){\circle*{2}}

\put(34,49){\makebox(0,0)[bl]{\(B^{1}_{1}=W_{1}\star D_{3}\star
O_{3}\)}}

\put(34,45){\makebox(0,0)[bl]{\(B^{1}_{2}=W_{2}\star D_{2}\star
O_{2}\)}}

\put(34,41){\makebox(0,0)[bl]{\(B^{1}_{3}=W_{1}\star D_{2}\star
O_{5}\)}}

\put(32,33){\line(0,1){20}}

\put(00,49){\line(0,1){05}} \put(00,49){\circle*{2}}
\put(00,42){\line(0,1){11}}

\put(02,49){\makebox(0,0)[bl]{\(A^{1}_{1}=M_{2}\star E_{2}\)}}

\put(00,38){\line(1,0){18}} \put(32,38){\line(1,0){26}}

\put(00,33){\line(0,1){10}} \put(18,33){\line(0,1){05}}
\put(32,33){\line(0,1){05}} \put(45,33){\line(0,1){05}}
\put(58,33){\line(0,1){05}}

\put(00,33){\circle*{1}}

\put(18,33){\circle*{1}} \put(32,33){\circle*{1}}
\put(45,33){\circle*{1}} \put(58,33){\circle*{1}}

\put(02,33){\makebox(0,0)[bl]{\(M\)}}
\put(14,33){\makebox(0,0)[bl]{\(E\)}}
\put(34,33){\makebox(0,0)[bl]{\(W\)}}
\put(47,33){\makebox(0,0)[bl]{\(D\)}}
\put(60,33){\makebox(0,0)[bl]{\(O\)}}

\put(53,29){\makebox(0,0)[bl]{Operation}}
\put(54,25){\makebox(0,0)[bl]{system}}
\put(56,21){\makebox(0,0)[bl]{\(O_{1}(3)\)}}
\put(56,17){\makebox(0,0)[bl]{\(O_{2}(2)\)}}
\put(56,13){\makebox(0,0)[bl]{\(O_{3}(1)\)}}
\put(56,9){\makebox(0,0)[bl]{\(O_{4}(3)\)}}
\put(56,05){\makebox(0,0)[bl]{\(O_{5}(1)\)}}

\put(41,29){\makebox(0,0)[bl]{DBMS}}
\put(42,25){\makebox(0,0)[bl]{\(D_{1}(3)\)}}
\put(42,21){\makebox(0,0)[bl]{\(D_{2}(1)\)}}
\put(42,17){\makebox(0,0)[bl]{\(D_{3}(2)\)}}

\put(30,29){\makebox(0,0)[bl]{Web}}
\put(30,25){\makebox(0,0)[bl]{server}}
\put(30,21){\makebox(0,0)[bl]{\(W_{1}(1)\)}}
\put(30,17){\makebox(0,0)[bl]{\(W_{2}(1)\)}}
\put(30,13){\makebox(0,0)[bl]{\(W_{3}(3)\)}}
\put(30,9){\makebox(0,0)[bl]{\(W_{4}(3)\)}}
\put(30,05){\makebox(0,0)[bl]{\(W_{5}(3)\)}}

\put(13,29){\makebox(0,0)[bl]{Server for}}
\put(13,25){\makebox(0,0)[bl]{applica-}}
\put(13,21){\makebox(0,0)[bl]{tions}}
\put(14,17){\makebox(0,0)[bl]{\(E_{1}(3)\)}}
\put(14,13){\makebox(0,0)[bl]{\(E_{2}(1)\)}}
\put(14,9){\makebox(0,0)[bl]{\(E_{3}(2)\)}}
\put(14,05){\makebox(0,0)[bl]{\(E_{4}(3)\)}}

\put(00,29){\makebox(0,0)[bl]{Server }}
\put(00,25){\makebox(0,0)[bl]{for DBs}}
\put(00,21){\makebox(0,0)[bl]{\(M_{1}(3)\)}}
\put(00,17){\makebox(0,0)[bl]{\(M_{2}(2)\)}}
\put(00,13){\makebox(0,0)[bl]{\(M_{3}(3)\)}}

\end{picture}
\end{center}

 Fig. 11 illustrates composite DAs for part \(B\).
 Clearly, the resultant composite DAs are the following:

 (1) \(S^{1}_{1}=A^{1}_{1}\star B^{1}_{1}=(M_{2}\star E_{2})\star (W_{1}\star
D_{3}\star O_{3})\);

 (2) \(S^{1}_{2}=A^{1}_{1}\star B^{1}_{2}=(M_{2}\star E_{2})\star (W_{2}\star
D_{2}\star O_{2})\);

 (3) \(S^{1}_{3}=A^{1}_{1}\star B^{1}_{3}=(M_{2}\star E_{2})\star (W_{1}\star
D_{2}\star O_{5})\).


\begin{center}
\begin{picture}(60,62)
\put(00,0){\makebox(0,0)[bl]{Fig. 10. Space of system quality for
\(A\)}}

\put(0,010){\line(0,1){40}} \put(0,010){\line(3,4){15}}
\put(0,050){\line(3,-4){15}}

\put(20,015){\line(0,1){40}} \put(20,015){\line(3,4){15}}
\put(20,055){\line(3,-4){15}}

\put(40,020){\line(0,1){40}} \put(40,020){\line(3,4){15}}
\put(40,060){\line(3,-4){15}}


\put(40,40){\circle*{2}}
\put(28.8,43){\makebox(0,0)[bl]{\(N(A^{1}_{1}),N(A^{2}_{1}),N(A^{2}_{2})\)}}

\put(40,60){\circle*{1}} \put(40,60){\circle{3}}


\put(23,58){\makebox(0,0)[bl]{The ideal}}
\put(27,55){\makebox(0,0)[bl]{point}}

\put(42,58){\makebox(0,0)[bl]{ \( N( A^{3}_{1} ) \) }}

\put(0,7){\makebox(0,0)[bl]{\(w=1\)}}
\put(20,12){\makebox(0,0)[bl]{\(w=2\)}}
\put(40,17){\makebox(0,0)[bl]{\(w=3\)}}
\end{picture}
\end{center}
\begin{center}
\begin{picture}(60,62)
\put(0,0){\makebox(0,0)[bl]{Fig. 11. Space of system quality for
\(B\)}}

\put(0,010){\line(0,1){40}} \put(0,010){\line(3,4){15}}
\put(0,050){\line(3,-4){15}}

\put(20,015){\line(0,1){40}} \put(20,015){\line(3,4){15}}
\put(20,055){\line(3,-4){15}}

\put(40,020){\line(0,1){40}} \put(40,020){\line(3,4){15}}
\put(40,060){\line(3,-4){15}}

\put(0,50){\circle*{2}}
\put(2.5,48){\makebox(0,0)[bl]{\(N(B^{1}_{3})\)}}

\put(42.5,48){\circle*{2}}
\put(35.5,42){\makebox(0,0)[bl]{\(N(B^{1}_{1}),N(B^{1}_{2})\)}}


\put(42.5,58){\makebox(0,0)[bl]{\(N(B^{2}_{1}),N(B^{2}_{2}),\)}}
\put(45.5,54){\makebox(0,0)[bl]{\(N(B^{3}_{1})\)}}

\put(40,60){\circle*{1}} \put(40,60){\circle{3}}


\put(23,58){\makebox(0,0)[bl]{The ideal}}
\put(27,55){\makebox(0,0)[bl]{point}}


\put(0,7){\makebox(0,0)[bl]{\(w=1\)}}
\put(20,12){\makebox(0,0)[bl]{\(w=2\)}}
\put(40,17){\makebox(0,0)[bl]{\(w=3\)}}
\end{picture}
\end{center}

 In addition,
 it is reasonable to consider the following
 technological system problems \cite{lev06}:
 (a) revelation of ``bottlenecks'' and
 (b) improvement of some obtained solution(s).
 For example, let us examine composite DAs
 for \(B\):
 ~\(B^{1}_{3} = W_{1}\star D_{2} \star O_{5}\)~
 with ~\(N(B^{1}_{3}) = (1;3,0,0)\).
 Here compatibility \(~(D_{2},O_{5})\) (that equals \(1\))
 is the ``bottleneck''.
 As a result, a special activity for improving this compatibility can
 be considered as an improvement operation.

\subsection{Corporate Application}

  The priorities of DAs for example 2 (corporate application)
  are shown in Fig. 12 in parentheses.

  For system part \(A\),
  we get the following Pareto-efficient composite DAs:
 ~\(A^{2}_{1} = M_{1} \star E_{1} \), \(N(A^{2}_{1})= (3;1,1,0)\); and
 ~\(A^{2}_{2} = M_{2} \star E_{2} \), \(N(A^{2}_{1})= (3;1,1,0)\).
%
 Quality of decisions
%
%
%
 ~\(A^{2}_{1} \) and  ~\(A^{2}_{2} \) is depicted in Fig. 10.
 For system part \(B\),
  we get the following Pareto-efficient composite DAs (the ideal solutions):
 ~\(B^{2}_{1} = W_{1} \star D_{3} \star O_{5}\), \(N(B^{2}_{1})=
 (3;3,0,0)\); and
 ~\(B^{2}_{2} = W_{2} \star D_{3} \star O_{2}\), \(N(B^{2}_{2})=
 (3;3,0,0)\).
 Quality of decisions
 ~\(B^{2}_{1} \) and  ~\(B^{2}_{2} \) is depicted in Fig. 11.
 As a result, we get the following four final composite DAs:

 (1) \(S^{2}_{1}=A^{2}_{1}\star B^{2}_{1}=(M_{1}\star E_{1})\star (W_{1}\star
D_{3}\star O_{5})\);

 (2) \(S^{2}_{2}=A^{2}_{1}\star B^{2}_{2}=(M_{1}\star E_{1})\star (W_{2}\star
D_{3}\star O_{2})\);

 (3) \(S^{2}_{3}=A^{2}_{2}\star B^{2}_{1}=(M_{2}\star E_{2})\star (W_{1}\star
D_{3}\star O_{5})\);
%

 (4) \(S^{2}_{4}=A^{2}_{2}\star B^{2}_{2}=(M_{2}\star E_{2})\star (W_{2}\star
D_{3}\star O_{2})\).

 Fig. 12 depicts information system and composite decisions for
 example 2.

\begin{center}
\begin{picture}(75,81)
\put(09,0){\makebox(0,0)[bl] {Fig. 12. Corporate application}}


\put(00,76){\circle*{3}}

\put(03,75){\makebox(0,0)[bl]{\(S=A\star B\)}}

\put(02,70){\makebox(0,0)[bl] {\(S^{2}_{1}=A^{2}_{1}\star
B^{2}_{1}=(M_{1}\star E_{1})\star (W_{1}\star D_{3}\star
O_{5})\)}}

\put(02,66){\makebox(0,0)[bl] {\(S^{2}_{2}=A^{2}_{1}\star
B^{2}_{2}=(M_{1}\star E_{1})\star (W_{2}\star D_{3}\star
O_{2})\)}}

\put(02,62){\makebox(0,0)[bl] {\(S^{2}_{3}=A^{2}_{1}\star
B^{2}_{1}=(M_{2}\star E_{2})\star (W_{1}\star D_{3}\star
O_{5})\)}}

\put(02,58){\makebox(0,0)[bl] {\(S^{2}_{4}=A^{2}_{1}\star
B^{2}_{2}=(M_{2}\star E_{2})\star (W_{2}\star D_{3}\star
O_{2})\)}}

\put(02,54){\makebox(0,0)[bl]{Hardware }}
\put(02,50){\makebox(0,0)[bl]{\(A=M\star E\)}}

\put(33,54){\makebox(0,0)[bl]{Software }}
\put(33,50){\makebox(0,0)[bl]{\(B=W\star D\star O\)}}

\put(00,48){\line(0,1){28}} \put(00,48){\line(1,0){32}}
\put(32,43){\line(0,1){05}} \put(32,43){\circle*{2}}

\put(34,43){\makebox(0,0)[bl]{\(B^{2}_{1}=W_{1}\star D_{3}\star
O_{5}\)}}

\put(34,39){\makebox(0,0)[bl]{\(B^{2}_{2}=W_{2}\star D_{3}\star
O_{2}\)}}

\put(32,33){\line(0,1){10}}

\put(00,43){\line(0,1){05}} \put(00,43){\circle*{2}}
\put(00,42){\line(0,1){11}}

\put(02,43){\makebox(0,0)[bl]{\(A^{2}_{1}=M_{1}\star E_{1}\)}}
\put(02,39){\makebox(0,0)[bl]{\(A^{2}_{2}=M_{2}\star E_{2}\)}}

\put(00,38){\line(1,0){18}} \put(32,38){\line(1,0){27}}

\put(00,33){\line(0,1){10}} \put(18,33){\line(0,1){05}}
\put(32,33){\line(0,1){05}} \put(45,33){\line(0,1){05}}
\put(59,33){\line(0,1){05}}

\put(00,33){\circle*{1}}

\put(18,33){\circle*{1}} \put(32,33){\circle*{1}}
\put(45,33){\circle*{1}} \put(59,33){\circle*{1}}

\put(02,33){\makebox(0,0)[bl]{\(M\)}}
\put(14,33){\makebox(0,0)[bl]{\(E\)}}
\put(33,33){\makebox(0,0)[bl]{\(W\)}}
\put(47,33){\makebox(0,0)[bl]{\(D\)}}
\put(61,33){\makebox(0,0)[bl]{\(O\)}}

\put(53,29){\makebox(0,0)[bl]{Operation}}
\put(54,25){\makebox(0,0)[bl]{system}}
\put(56,21){\makebox(0,0)[bl]{\(O_{1}(3)\)}}
\put(56,17){\makebox(0,0)[bl]{\(O_{2}(1)\)}}
\put(56,13){\makebox(0,0)[bl]{\(O_{3}(2)\)}}
\put(56,9){\makebox(0,0)[bl]{\(O_{4}(2)\)}}
\put(56,05){\makebox(0,0)[bl]{\(O_{5}(1)\)}}

\put(41,29){\makebox(0,0)[bl]{DBMS}}
\put(42,25){\makebox(0,0)[bl]{\(D_{1}(3)\)}}
\put(42,21){\makebox(0,0)[bl]{\(D_{2}(2)\)}}
\put(42,17){\makebox(0,0)[bl]{\(D_{3}(1)\)}}

\put(30,29){\makebox(0,0)[bl]{Web}}
\put(30,25){\makebox(0,0)[bl]{server}}
\put(30,21){\makebox(0,0)[bl]{\(W_{1}(1)\)}}
\put(30,17){\makebox(0,0)[bl]{\(W_{2}(2)\)}}
\put(30,13){\makebox(0,0)[bl]{\(W_{3}(3)\)}}
\put(30,9){\makebox(0,0)[bl]{\(W_{4}(3)\)}}
\put(30,05){\makebox(0,0)[bl]{\(W_{5}(3)\)}}

\put(13,29){\makebox(0,0)[bl]{Server for}}
\put(13,25){\makebox(0,0)[bl]{applica-}}
\put(13,21){\makebox(0,0)[bl]{tions}}
\put(14,17){\makebox(0,0)[bl]{\(E_{1}(1)\)}}
\put(14,13){\makebox(0,0)[bl]{\(E_{2}(3)\)}}
\put(14,9){\makebox(0,0)[bl]{\(E_{3}(3)\)}}
\put(14,05){\makebox(0,0)[bl]{\(E_{4}(3)\)}}

\put(00,29){\makebox(0,0)[bl]{Server }}
\put(00,25){\makebox(0,0)[bl]{for DBs}}
\put(00,21){\makebox(0,0)[bl]{\(M_{1}(2)\)}}
\put(00,17){\makebox(0,0)[bl]{\(M_{2}(2)\)}}
\put(00,13){\makebox(0,0)[bl]{\(M_{3}(3)\)}}

\end{picture}
\end{center}

\subsection{Academic Application}

 The priorities of DAs for example 3 (academic application)
  are shown in Fig. 13 in parentheses.
 For system part \(A\),
 we get the following Pareto-efficient composite DA:
 ~\(A^{3}_{1} = M_{3} \star E_{2} \), \(N(A^{3}_{1})= (3;2,0,0)\).
 Quality of decision  ~\(A^{3}_{1} \) is depicted in Fig. 10.
 For system part \(B\),
  we get the following Pareto-efficient composite DA:
 ~\(B^{3}_{1} = W_{1} \star D_{2} \star O_{3}\), \(N(B^{3}_{1})=
 (3;3,0,0)\).
%
%
 Quality of decision  ~\(B^{3}_{1} \) is depicted in Fig. 11.
 The resultant composite DA is the following:

 \(S^{3}_{1}=A^{3}_{1}\star B^{3}_{1}=(M_{3}\star E_{2})\star (W_{1}\star
D_{2}\star O_{3})\).

 Fig. 13 depicts information system and composite decisions for
 example 3.

\subsection{Towards Analysis of Decisions}

 Table 7 summarizes the resultant composite decisions
 for three considered applied examples above
 and it is a basis to analyze and/or compare the corresponding resultant decisions.

 Evidently, certain requirements and constraints lead to
 specific results. For example,
  in the corporate applications
 maintenance requirements can be important,
 in the academic applications
 performance requirements can be often crucial ones.
 In the article, numerical results have only illustrative
 character to explain the methodological approach
 (i.e., steps of solving scheme).
 The design and usage of special approaches to analysis and
 comparison of different resultant  applied decisions
 is a prospective topic for future studies
 (e.g., multicriteria comparison, stability analysis).

\begin{center}
\begin{picture}(75,72)
\put(08,01){\makebox(0,0)[bl] {Fig. 13. Academic application}}


\put(00,67){\circle*{3}}

\put(03,66){\makebox(0,0)[bl]{\(S=A\star B\)}}

\put(02,61){\makebox(0,0)[bl] {\(S^{3}_{1}=A^{3}_{1}\star
B^{3}_{1}=(M_{3}\star E_{2})\star (W_{1}\star D_{2}\star
O_{3})\)}}

\put(02,56){\makebox(0,0)[bl]{Hardware}}
\put(02,52){\makebox(0,0)[bl]{\(A=M\star E\)}}

\put(33,56){\makebox(0,0)[bl]{Software}}
\put(33,52){\makebox(0,0)[bl]{\(B=W\star D\star O\)}}

\put(00,50){\line(0,1){17}} \put(00,50){\line(1,0){32}}
\put(32,45){\line(0,1){05}} \put(32,45){\circle*{2}}

\put(34,45){\makebox(0,0)[bl]{\(B^{3}_{1}=W_{1}\star D_{2}\star
O_{3}\)}}

\put(32,35){\line(0,1){10}}

\put(00,45){\line(0,1){05}} \put(00,45){\circle*{2}}
\put(00,44){\line(0,1){11}}

\put(02,45){\makebox(0,0)[bl]{\(A^{3}_{1}=M_{3}\star E_{2}\)}}


\put(00,40){\line(1,0){18}} \put(32,40){\line(1,0){27}}

\put(00,35){\line(0,1){10}} \put(18,35){\line(0,1){05}}
\put(32,35){\line(0,1){05}} \put(45,35){\line(0,1){05}}
\put(59,35){\line(0,1){05}}

\put(00,35){\circle*{1}} \put(18,35){\circle*{1}}
\put(32,35){\circle*{1}} \put(45,35){\circle*{1}}
\put(59,35){\circle*{1}}

\put(02,35){\makebox(0,0)[bl]{\(M\)}}
\put(14,35){\makebox(0,0)[bl]{\(E\)}}
\put(34,35){\makebox(0,0)[bl]{\(W\)}}
\put(47,35){\makebox(0,0)[bl]{\(D\)}}
\put(61,35){\makebox(0,0)[bl]{\(O\)}}

\put(53,31){\makebox(0,0)[bl]{Operation}}
\put(54,27){\makebox(0,0)[bl]{system}}
\put(56,23){\makebox(0,0)[bl]{\(O_{1}(3)\)}}
\put(56,19){\makebox(0,0)[bl]{\(O_{2}(2)\)}}
\put(56,15){\makebox(0,0)[bl]{\(O_{3}(1)\)}}
\put(56,11){\makebox(0,0)[bl]{\(O_{4}(2)\)}}
\put(56,07){\makebox(0,0)[bl]{\(O_{5}(1)\)}}

\put(41,31){\makebox(0,0)[bl]{DBMS}}
\put(42,27){\makebox(0,0)[bl]{\(D_{1}(3)\)}}
\put(42,23){\makebox(0,0)[bl]{\(D_{2}(1)\)}}
\put(42,19){\makebox(0,0)[bl]{\(D_{3}(2)\)}}

\put(30,31){\makebox(0,0)[bl]{Web}}
\put(30,27){\makebox(0,0)[bl]{server}}
\put(30,23){\makebox(0,0)[bl]{\(W_{1}(1)\)}}
\put(30,19){\makebox(0,0)[bl]{\(W_{2}(2)\)}}
\put(30,15){\makebox(0,0)[bl]{\(W_{3}(3)\)}}
\put(30,11){\makebox(0,0)[bl]{\(W_{4}(3)\)}}
\put(30,07){\makebox(0,0)[bl]{\(W_{5}(3)\)}}

\put(13,31){\makebox(0,0)[bl]{Server for}}
\put(13,27){\makebox(0,0)[bl]{applica-}}
\put(13,23){\makebox(0,0)[bl]{tions}}
\put(14,19){\makebox(0,0)[bl]{\(E_{1}(3)\)}}
\put(14,15){\makebox(0,0)[bl]{\(E_{2}(1)\)}}
\put(14,11){\makebox(0,0)[bl]{\(E_{3}(2)\)}}
\put(14,07){\makebox(0,0)[bl]{\(E_{4}(3)\)}}

\put(00,31){\makebox(0,0)[bl]{Server }}
\put(00,27){\makebox(0,0)[bl]{for DBs}}
\put(00,23){\makebox(0,0)[bl]{\(M_{1}(3)\)}}
\put(00,19){\makebox(0,0)[bl]{\(M_{2}(2)\)}}
\put(00,15){\makebox(0,0)[bl]{\(M_{3}(1)\)}}

\end{picture}
\end{center}

\begin{center}
\begin{picture}(73,53)
\put(07,49){\makebox(0,0)[bl]{Table 7. Resultant composite
decisions}}

\put(00,0){\line(1,0){73}}

 \put(00,41){\line(1,0){73}}
\put(00,47){\line(1,0){73}}

\put(00,0){\line(0,1){47}} \put(04,00){\line(0,1){47}}
\put(73,0){\line(0,1){47}}

\put(0.5,43){\makebox(0,0)[bl]{\#}}

\put(25,43){\makebox(0,0)[bl]{Composite DAs}}


\put(0.5,36.5){\makebox(0,0)[bl]{1.}}
\put(0.5,21.5){\makebox(0,0)[bl]{2.}}
\put(0.5,1.5){\makebox(0,0)[bl]{3.}}

\put(05,36){\makebox(0,0)[bl] {\(S^{1}_{1}=A^{1}_{1}\star
B^{1}_{1}=(M_{2}\star E_{2})\star (W_{1}\star D_{3}\star
O_{3})\)}}

\put(05,31){\makebox(0,0)[bl] {\(S^{1}_{2}=A^{1}_{1}\star
B^{1}_{2}=(M_{2}\star E_{2})\star (W_{5}\star D_{2}\star
O_{2})\)}}

\put(05,26){\makebox(0,0)[bl] {\(S^{1}_{3}=A^{1}_{1}\star
B^{1}_{3}=(M_{2}\star E_{2})\star (W_{1}\star D_{2}\star
O_{5})\)}}


\put(05,21){\makebox(0,0)[bl] {\(S^{2}_{1}=A^{2}_{1}\star
B^{2}_{1}=(M_{1}\star E_{1})\star (W_{1}\star D_{3}\star
O_{5})\)}}

\put(05,16){\makebox(0,0)[bl] {\(S^{2}_{2}=A^{2}_{1}\star
B^{2}_{2}=(M_{1}\star E_{1})\star (W_{2}\star D_{3}\star
O_{2})\)}}

\put(05,11){\makebox(0,0)[bl] {\(S^{2}_{3}=A^{2}_{1}\star
B^{2}_{1}=(M_{2}\star E_{2})\star (W_{1}\star D_{3}\star
O_{5})\)}}

\put(05,6){\makebox(0,0)[bl] {\(S^{2}_{4}=A^{2}_{1}\star
B^{2}_{2}=(M_{2}\star E_{2})\star (W_{2}\star D_{3}\star
O_{2})\)}}


\put(05,01){\makebox(0,0)[bl] {\(S^{3}_{1}=A^{3}_{1}\star
B^{3}_{1}=(M_{3}\star E_{2})\star (W_{1}\star D_{2}\star
O_{3})\)}}

\end{picture}
\end{center}

\subsection{Usage of Multiple Choice Problem}

 In this case estimates of compatibility are not used and the
 model is more simple.
 Here we consider the greedy heuristic for applied example 1
 (communication service provider).
 Let us compute for each DA(\(\mu \))
 a priority
  ~\(\overline{r}(\mu )\)
 by three criteria ~\(C_{2}\),
 \(C_{3}\), and \(C_{4}\).
 After that it is possible to get for each DA the value
 (as ``relative utility'')
 ~\(\lambda(\mu )  =  ( \widehat{r}   -       \overline{r}(\mu ) )    / z_{\mu }\)
 (where \(\widehat{r} = \max_{\mu} \{ \overline{r}(\mu ) \}\)
 and
 \(z_{\mu }\) is the estimate of cost for DA(\(\mu \))
  by criterion \(C_{1}\)).
 As a result, we can get a linear ordering of all DAs by
 ~\(\lambda(\mu )\) to get the number of linear order ~\(\pi(\mu )\).
 Tables 8 and 9 contains estimates
~\(\overline{r}(\mu ) \), ~\(\lambda(\mu )\), and ~\(\pi(\mu )\).

\begin{center}
\begin{picture}(40,62)
\put(06,58){\makebox(0,0)[bl]{Table 8. Ordering}}

\put(00,0){\line(1,0){40}} \put(00,50){\line(1,0){40}}
\put(00,56){\line(1,0){40}}

\put(00,0){\line(0,1){56}} \put(10,00){\line(0,1){56}}
\put(40,0){\line(0,1){56}}

\put(20,50){\line(0,1){6}} \put(30,50){\line(0,1){6}}

\put(01,52){\makebox(0,0)[bl]{DAs}}

\put(12,52){\makebox(0,0)[bl]{\( \overline{r}(\mu) \)}}
\put(22,52){\makebox(0,0)[bl]{\(\lambda(\mu)\)}}
\put(32,52){\makebox(0,0)[bl]{\(\pi(\mu)\)}}


\put(01,46){\makebox(0,0)[bl]{\(M_{1}\)}}
\put(01,42){\makebox(0,0)[bl]{\(M_{2}\)}}
\put(01,38){\makebox(0,0)[bl]{\(M_{3}\)}}
\put(01,34){\makebox(0,0)[bl]{\(E_{1}\)}}
\put(01,30){\makebox(0,0)[bl]{\(E_{2}\)}}
\put(01,26){\makebox(0,0)[bl]{\(E_{3}\)}}
\put(01,22){\makebox(0,0)[bl]{\(E_{4}\)}}
\put(01,18){\makebox(0,0)[bl]{\(W_{1}\)}}
\put(01,14){\makebox(0,0)[bl]{\(W_{2}\)}}
\put(01,10){\makebox(0,0)[bl]{\(W_{3}\)}}
\put(01,06){\makebox(0,0)[bl]{\(W_{4}\)}}
\put(01,02){\makebox(0,0)[bl]{\(W_{5}\)}}

\put(15,46){\makebox(0,0)[bl]{\(3\)}}
\put(22,46){\makebox(0,0)[bl]{\(0.00\)}}
\put(34,46){\makebox(0,0)[bl]{\(13\)}}

\put(15,42){\makebox(0,0)[bl]{\(2\)}}
\put(22,42){\makebox(0,0)[bl]{\(0.20\)}}
\put(34,42){\makebox(0,0)[bl]{\(11\)}}

\put(15,38){\makebox(0,0)[bl]{\(1\)}}
\put(22,38){\makebox(0,0)[bl]{\(0.33\)}}
\put(35,38){\makebox(0,0)[bl]{\(7\)}}

\put(15,34){\makebox(0,0)[bl]{\(3\)}}
\put(22,34){\makebox(0,0)[bl]{\(0.00\)}}
\put(34,34){\makebox(0,0)[bl]{\(14\)}}

\put(15,30){\makebox(0,0)[bl]{\(1\)}}
\put(22,30){\makebox(0,0)[bl]{\(0.33\)}}
\put(35,30){\makebox(0,0)[bl]{\(8\)}}

\put(15,26){\makebox(0,0)[bl]{\(2\)}}
\put(22,26){\makebox(0,0)[bl]{\(0.20\)}}
\put(34,26){\makebox(0,0)[bl]{\(12\)}}

\put(15,22){\makebox(0,0)[bl]{\(3\)}}
\put(22,22){\makebox(0,0)[bl]{\(0.00\)}}
\put(34,22){\makebox(0,0)[bl]{\(15\)}}

\put(15,18){\makebox(0,0)[bl]{\(1\)}}
\put(22,18){\makebox(0,0)[bl]{\(2.00\)}}
\put(35,18){\makebox(0,0)[bl]{\(1\)}}

\put(15,14){\makebox(0,0)[bl]{\(2\)}}
\put(22,14){\makebox(0,0)[bl]{\(0.25\)}}
\put(34,14){\makebox(0,0)[bl]{\(10\)}}

\put(15,10){\makebox(0,0)[bl]{\(3\)}}
\put(22,10){\makebox(0,0)[bl]{\(0.00\)}}
\put(34,10){\makebox(0,0)[bl]{\(16\)}}

\put(15,06){\makebox(0,0)[bl]{\(3\)}}
\put(22,06){\makebox(0,0)[bl]{\(0.00\)}}
\put(34,06){\makebox(0,0)[bl]{\(17\)}}

\put(15,02){\makebox(0,0)[bl]{\(3\)}}
\put(22,02){\makebox(0,0)[bl]{\(0.00\)}}
\put(34,02){\makebox(0,0)[bl]{\(18\)}}

\end{picture}
%
\begin{picture}(40,46)
\put(06,42){\makebox(0,0)[bl]{Table 9. Ordering }}

\put(00,0){\line(1,0){40}} \put(00,34){\line(1,0){40}}
\put(00,40){\line(1,0){40}}

\put(00,0){\line(0,1){40}} \put(10,00){\line(0,1){40}}
\put(40,0){\line(0,1){40}}

\put(20,34){\line(0,1){6}} \put(30,34){\line(0,1){6}}

\put(01,36){\makebox(0,0)[bl]{DAs}}

\put(12,36){\makebox(0,0)[bl]{\( \overline{r}(\mu) \)}}
\put(22,36){\makebox(0,0)[bl]{\(\lambda(\mu)\)}}
\put(32,36){\makebox(0,0)[bl]{\(\pi(\mu)\)}}


\put(01,30){\makebox(0,0)[bl]{\(D_{1}\)}}
\put(01,26){\makebox(0,0)[bl]{\(D_{2}\)}}
\put(01,22){\makebox(0,0)[bl]{\(D_{3}\)}}
\put(01,18){\makebox(0,0)[bl]{\(O_{1}\)}}
\put(01,14){\makebox(0,0)[bl]{\(O_{2}\)}}
\put(01,10){\makebox(0,0)[bl]{\(O_{3}\)}}
\put(01,06){\makebox(0,0)[bl]{\(O_{4}\)}}
\put(01,02){\makebox(0,0)[bl]{\(O_{5}\)}}

\put(15,30){\makebox(0,0)[bl]{\(1\)}}
\put(22,30){\makebox(0,0)[bl]{\(0.33\)}}
\put(35,30){\makebox(0,0)[bl]{\(9\)}}

\put(15,26){\makebox(0,0)[bl]{\(1\)}}
\put(22,26){\makebox(0,0)[bl]{\(0.40\)}}
\put(35,26){\makebox(0,0)[bl]{\(6\)}}

\put(15,22){\makebox(0,0)[bl]{\(3\)}}
\put(22,22){\makebox(0,0)[bl]{\(0.00\)}}
\put(34,22){\makebox(0,0)[bl]{\(19\)}}

\put(15,18){\makebox(0,0)[bl]{\(3\)}}
\put(22,18){\makebox(0,0)[bl]{\(0.00\)}}
\put(34,18){\makebox(0,0)[bl]{\(20\)}}

\put(15,14){\makebox(0,0)[bl]{\(3\)}}
\put(22,14){\makebox(0,0)[bl]{\(0.50\)}}
\put(35,14){\makebox(0,0)[bl]{\(5\)}}

\put(15,10){\makebox(0,0)[bl]{\(1\)}}
\put(22,10){\makebox(0,0)[bl]{\(1.00\)}}
\put(35,10){\makebox(0,0)[bl]{\(2\)}}

\put(15,06){\makebox(0,0)[bl]{\(2\)}}
\put(22,06){\makebox(0,0)[bl]{\(1.00\)}}
\put(35,06){\makebox(0,0)[bl]{\(3\)}}

\put(15,02){\makebox(0,0)[bl]{\(2\)}}
\put(22,02){\makebox(0,0)[bl]{\(1.00\)}}
\put(35,02){\makebox(0,0)[bl]{\(4\)}}

\end{picture}
\end{center}

 As a result, the following solutions are obtained:
 ~(1) total cost constraint \( \leq 15  \):
 ~\( \widetilde{S}^{1}_{1} = M_{1} \star E_{2} \star W_{1} \star D_{2} \star O_{3}
 \);
 ~(2) total cost constraint \( \leq 18  \):
 ~\( \widetilde{S}^{1}_{2} = M_{2} \star E_{2} \star W_{1} \star D_{2} \star O_{3}
 \); and
  ~(3) total cost constraint \( \leq 19  \):
 ~\( \widetilde{S}^{1}_{3} = M_{3} \star E_{2} \star W_{1} \star D_{2} \star O_{3}
 \).

\subsection{Design of System Trajectory}

 The scheme of multistage
 design consists of two phases (Fig. 14):
 ~{\it 1.} design of composite DAs for each time stage
 (HMMD);
 ~{\it 2.} design of a system trajectory based on DAs
 which were obtained at phase 1 (HMMD).
 Note a change of elements into the trajectory can require
 some efforts, and it is necessary to solve an additional top-level
 composition problem (phase 2) as follows:

  {\it Combine a trajectory
  (i.e., selection of a system solution at each stage )
  while taking into account quality of composite DAs at each stage
  and a cost of the component changes.}

\begin{center}
\begin{picture}(70,49)
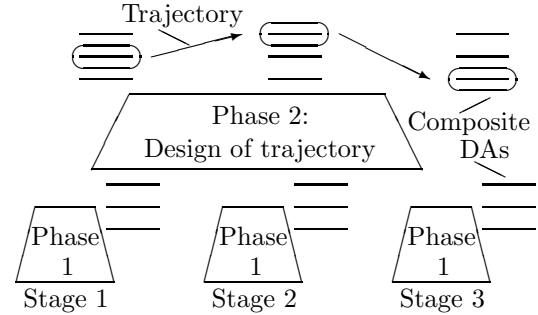
\put(04,01){\makebox(0,0)[bl]{Fig. 14. Illustration of multistage
design}}


\put(0,10){\line(1,0){13}} \put(0,10){\line(1,4){2.5}}
\put(13,10){\line(-1,4){2.5}} \put(2.5,20){\line(1,0){8}}

\put(2,15){\makebox(0,0)[bl]{Phase}}
\put(6,11){\makebox(0,0)[bl]{1}}


\put(12,23){\line(1,0){7}} \put(12,20){\line(1,0){7}}
\put(12,17){\line(1,0){7}}

\put(01,06){\makebox(0,0)[bl]{Stage 1}}


\put(25,10){\line(1,0){13}} \put(25,10){\line(1,4){2.5}}
\put(38,10){\line(-1,4){2.5}} \put(27.5,20){\line(1,0){8}}

\put(27,15){\makebox(0,0)[bl]{Phase}}
\put(31,11){\makebox(0,0)[bl]{1}}


\put(37,23){\line(1,0){7}} \put(37,20){\line(1,0){7}}
\put(37,17){\line(1,0){7}}

\put(26,06){\makebox(0,0)[bl]{Stage 2}}

\put(50,10){\line(1,0){13}} \put(50,10){\line(1,4){2.5}}
\put(63,10){\line(-1,4){2.5}} \put(52.5,20){\line(1,0){8}}

\put(52,15){\makebox(0,0)[bl]{Phase}}
\put(56,11){\makebox(0,0)[bl]{1}}


\put(62,23){\line(1,0){7}} \put(62,20){\line(1,0){7}}
\put(62,17){\line(1,0){7}}

\put(51,06){\makebox(0,0)[bl]{Stage 3}}

\put(52,29.5){\makebox(0,0)[bl]{Composite}}
\put(59,26.5){\makebox(0,0)[bl]{DAs}}

\put(59,32.5){\line(2,1){4}}\put(61,26){\line(2,-1){4}}


\put(10,25){\line(1,0){44}}

\put(10,25){\line(1,2){5}} \put(54,25){\line(-1,2){5}}

\put(15,35){\line(1,0){34}}

\put(26,31){\makebox(0,0)[bl]{Phase 2:}}
\put(17,26){\makebox(0,0)[bl]{Design of trajectory}}


\put(14,44){\makebox(0,0)[bl]{Trajectory}}
\put(19,43.5){\line(2,-1){4}}

\put(12,40){\oval(9,3)}

\put(8.5,43){\line(1,0){7}} \put(8.5,40){\line(1,0){7}}
\put(8.5,37){\line(1,0){7}}

\put(18,40){\vector(4,1){12}}


\put(37,43){\oval(9,3)}

\put(33.5,43){\line(1,0){7}} \put(33.5,40){\line(1,0){7}}
\put(33.5,37){\line(1,0){7}}

\put(43,43){\vector(2,-1){12}}



\put(62,37){\oval(9,3)}

\put(58.5,43){\line(1,0){7}} \put(58.5,40){\line(1,0){7}}
\put(58.5,37){\line(1,0){7}}

\end{picture}
\end{center}

\begin{center}
\begin{picture}(75,71)
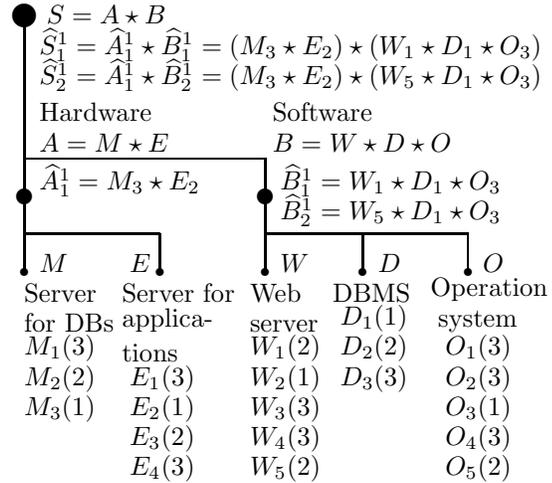
\put(01,0){\makebox(0,0)[bl] {Fig. 15. Communication provider
(stage 2)}}


\put(00,67){\circle*{3}}

\put(03,66){\makebox(0,0)[bl]{\(S=A\star B\)}}

\put(02,61){\makebox(0,0)[bl] {\( \widehat{S}^{1}_{1} =
 \widehat{A}^{1}_{1} \star \widehat{B}^{1}_{1}=(M_{3}\star E_{2})\star (W_{1}\star
D_{1}\star O_{3})\)}}

\put(02,57){\makebox(0,0)[bl] {\( \widehat{S}^{1}_{2} =
\widehat{A}^{1}_{1}\star \widehat{B}^{1}_{2} = ( M_{3}\star E_{2}
)\star ( W_{5} \star D_{1} \star O_{3} ) \)}}

\put(02,53){\makebox(0,0)[bl]{Hardware }}
\put(02,49){\makebox(0,0)[bl]{\(A=M\star E\)}}

\put(33,53){\makebox(0,0)[bl]{Software }}
\put(33,49){\makebox(0,0)[bl]{\(B=W\star D\star O\)}}

\put(00,48){\line(0,1){20}} \put(00,48){\line(1,0){32}}
\put(32,43){\line(0,1){05}}

\put(32,43){\circle*{2}}

\put(34,43){\makebox(0,0)[bl]{\( \widehat{B}^{1}_{1}=W_{1}\star
D_{1}\star O_{3}\)}}

\put(34,39){\makebox(0,0)[bl]{\( \widehat{B}^{1}_{2}=W_{5}\star
D_{1}\star O_{3}\)}}

\put(32,33){\line(0,1){10}}

\put(00,43){\line(0,1){05}} \put(00,43){\circle*{2}}
\put(00,42){\line(0,1){11}}

\put(02,43){\makebox(0,0)[bl]{\( \widehat{A}^{1}_{1}=M_{3}\star
E_{2}\)}}

\put(00,38){\line(1,0){18}} \put(32,38){\line(1,0){27}}

\put(00,33){\line(0,1){10}} \put(18,33){\line(0,1){05}}
\put(32,33){\line(0,1){05}} \put(45,33){\line(0,1){05}}
\put(59,33){\line(0,1){05}} \put(00,33){\circle*{1}}

\put(18,33){\circle*{1}} \put(32,33){\circle*{1}}
\put(45,33){\circle*{1}} \put(59,33){\circle*{1}}

\put(02,33){\makebox(0,0)[bl]{\(M\)}}
\put(14,33){\makebox(0,0)[bl]{\(E\)}}
\put(34,33){\makebox(0,0)[bl]{\(W\)}}
\put(47,33){\makebox(0,0)[bl]{\(D\)}}
\put(61,33){\makebox(0,0)[bl]{\(O\)}}

\put(54,29){\makebox(0,0)[bl]{Operation}}
\put(55,25){\makebox(0,0)[bl]{system}}
\put(56,21){\makebox(0,0)[bl]{\(O_{1}(3)\)}}
\put(56,17){\makebox(0,0)[bl]{\(O_{2}(3)\)}}
\put(56,13){\makebox(0,0)[bl]{\(O_{3}(1)\)}}
\put(56,9){\makebox(0,0)[bl]{\(O_{4}(3)\)}}
\put(56,05){\makebox(0,0)[bl]{\(O_{5}(2)\)}}

\put(41,29){\makebox(0,0)[bl]{DBMS}}
\put(42,25){\makebox(0,0)[bl]{\(D_{1}(1)\)}}
\put(42,21){\makebox(0,0)[bl]{\(D_{2}(2)\)}}
\put(42,17){\makebox(0,0)[bl]{\(D_{3}(3)\)}}

\put(30,29){\makebox(0,0)[bl]{Web}}
\put(30,25){\makebox(0,0)[bl]{server}}
\put(30,21){\makebox(0,0)[bl]{\(W_{1}(2)\)}}
\put(30,17){\makebox(0,0)[bl]{\(W_{2}(1)\)}}
\put(30,13){\makebox(0,0)[bl]{\(W_{3}(3)\)}}
\put(30,9){\makebox(0,0)[bl]{\(W_{4}(3)\)}}
\put(30,05){\makebox(0,0)[bl]{\(W_{5}(2)\)}}

\put(13,29){\makebox(0,0)[bl]{Server for}}
\put(13,25){\makebox(0,0)[bl]{applica-}}
\put(13,21){\makebox(0,0)[bl]{tions}}
\put(14,17){\makebox(0,0)[bl]{\(E_{1}(3)\)}}
\put(14,13){\makebox(0,0)[bl]{\(E_{2}(1)\)}}
\put(14,9){\makebox(0,0)[bl]{\(E_{3}(2)\)}}
\put(14,05){\makebox(0,0)[bl]{\(E_{4}(3)\)}}

\put(00,29){\makebox(0,0)[bl]{Server }}
\put(00,25){\makebox(0,0)[bl]{for DBs}}
\put(00,21){\makebox(0,0)[bl]{\(M_{1}(3)\)}}
\put(00,17){\makebox(0,0)[bl]{\(M_{2}(2)\)}}
\put(00,13){\makebox(0,0)[bl]{\(M_{3}(1)\)}}

\end{picture}
\end{center}

 This problem (trajectory design) is presented in
  (\cite{lev05}, \cite{lev06}).
 Here example 1 (communication service provider) is considered
 for three stages.
 Stage 1 corresponds to Fig. 9 with solutions
 ~\(S^{1}_{1}\), \(S^{1}_{2}\), and \(S^{1}_{3}\).

 For stage 2 (near future) and stage 3 (future)
 other weights of criteria are used:
 ~stage 2: ~\(-1\), \(3\), \(-1\), and \(3\);
 ~stage 3: ~\(-1\), \(5\), \(-3\), and \(5\).
 Fig. 15 and 16 depict results for stages 2 and 3.

 The composite DAs for stage 2 are
 the following:

 \( \widehat{A}^{1}_{1} = M_{3} \star E_{2} \), ~\( N ( \widehat{A}^{1}_{1}) = (3;2,0,0) \);

 \( \widehat{B}^{1}_{1} = W_{1} \star D_{1} \star O_{3} \), ~\( N( \widehat{B}^{1}_{1}) = (3;2,1,0) \);

 \( \widehat{B}^{1}_{2} = W_{5} \star D_{1} \star O_{3} \), ~\( N( \widehat{B}^{1}_{2}) = (3;2,1,0) \);

 \( \widehat{S}^{1}_{1} = ( \widehat{A}^{1}_{1} \star
  \widehat{B}^{1}_{1}) =
   ( M_{3} \star E_{2} ) \star  ( W_{1} \star D_{1} \star O_{3} )\);

 \( \widehat{S}^{1}_{2} = ( \widehat{A}^{1}_{1} \star \widehat{B}^{1}_{2}) =
   ( M_{3} \star E_{2} ) \star  ( W_{5} \star D_{1} \star O_{3} )\).

 The composite DAs for stage 3 are
 the following:

 \( \overline{A}^{1}_{1} = M_{3} \star E_{2} \), ~\( N ( \overline{A}^{1}_{1}) = (3;2,0,0) \);

 \( \overline{B}^{1}_{1} = W_{2} \star D_{2} \star O_{2} \), ~\( N( \overline{B}^{1}_{1}) = (3;2,1,0) \);

 \( \overline{S}^{1}_{1} = ( \overline{A}^{1}_{1} \star \overline{B}^{1}_{1}) =
 ( M_{3} \star E_{2} ) \star  ( W_{2} \star D_{2} \star O_{2} )\).

\begin{center}
\begin{picture}(75,68)
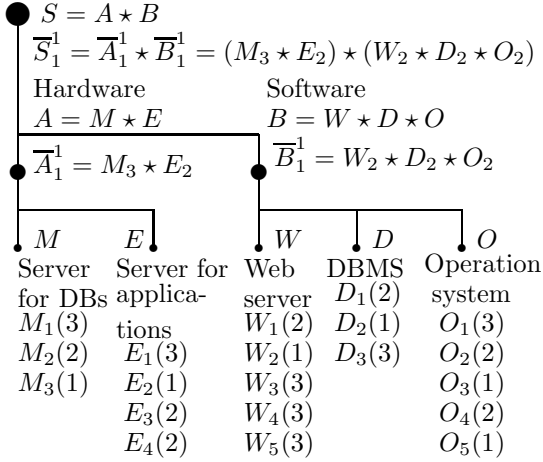
\put(01,0){\makebox(0,0)[bl] {Fig. 16. Communication provider
(stage 3)}}


\put(00,64){\circle*{3}}

\put(03,63){\makebox(0,0)[bl]{\(S=A\star B\)}}

\put(02,57){\makebox(0,0)[bl] {\( \overline{S}^{1}_{1} =
\overline{A}^{1}_{1}\star \overline{B}^{1}_{1}=(M_{3}\star
E_{2})\star (W_{2}\star D_{2}\star O_{2})\)}}

\put(02,53){\makebox(0,0)[bl]{Hardware }}
\put(02,49){\makebox(0,0)[bl]{\(A=M\star E\)}}

\put(33,53){\makebox(0,0)[bl]{Software }}
\put(33,49){\makebox(0,0)[bl]{\(B=W\star D\star O\)}}

\put(00,48){\line(0,1){16}} \put(00,48){\line(1,0){32}}
\put(32,43){\line(0,1){05}} \put(32,43){\circle*{2}}

\put(34,43){\makebox(0,0)[bl]{\( \overline{B}^{1}_{1}=W_{2}\star
D_{2}\star O_{2}\)}}

\put(32,33){\line(0,1){10}}

\put(00,43){\line(0,1){05}} \put(00,43){\circle*{2}}
\put(00,42){\line(0,1){11}}

\put(02,42){\makebox(0,0)[bl]{\( \overline{A}^{1}_{1}=M_{3}\star
E_{2}\)}}

\put(00,38){\line(1,0){18}} \put(32,38){\line(1,0){27}}

\put(00,33){\line(0,1){10}} \put(18,33){\line(0,1){05}}
\put(32,33){\line(0,1){05}} \put(45,33){\line(0,1){05}}
\put(59,33){\line(0,1){05}} \put(00,33){\circle*{1}}

\put(18,33){\circle*{1}} \put(32,33){\circle*{1}}
\put(45,33){\circle*{1}} \put(59,33){\circle*{1}}

\put(02,33){\makebox(0,0)[bl]{\(M\)}}
\put(14,33){\makebox(0,0)[bl]{\(E\)}}
\put(34,33){\makebox(0,0)[bl]{\(W\)}}
\put(47,33){\makebox(0,0)[bl]{\(D\)}}
\put(61,33){\makebox(0,0)[bl]{\(O\)}}

\put(54,29){\makebox(0,0)[bl]{Operation}}
\put(55,25){\makebox(0,0)[bl]{system}}
\put(56,21){\makebox(0,0)[bl]{\(O_{1}(3)\)}}
\put(56,17){\makebox(0,0)[bl]{\(O_{2}(2)\)}}
\put(56,13){\makebox(0,0)[bl]{\(O_{3}(1)\)}}
\put(56,9){\makebox(0,0)[bl]{\(O_{4}(2)\)}}
\put(56,05){\makebox(0,0)[bl]{\(O_{5}(1)\)}}

\put(41,29){\makebox(0,0)[bl]{DBMS}}
\put(42,25){\makebox(0,0)[bl]{\(D_{1}(2)\)}}
\put(42,21){\makebox(0,0)[bl]{\(D_{2}(1)\)}}
\put(42,17){\makebox(0,0)[bl]{\(D_{3}(3)\)}}

\put(30,29){\makebox(0,0)[bl]{Web}}
\put(30,25){\makebox(0,0)[bl]{server}}
\put(30,21){\makebox(0,0)[bl]{\(W_{1}(2)\)}}
\put(30,17){\makebox(0,0)[bl]{\(W_{2}(1)\)}}
\put(30,13){\makebox(0,0)[bl]{\(W_{3}(3)\)}}
\put(30,9){\makebox(0,0)[bl]{\(W_{4}(3)\)}}
\put(30,05){\makebox(0,0)[bl]{\(W_{5}(3)\)}}

\put(13,29){\makebox(0,0)[bl]{Server for}}
\put(13,25){\makebox(0,0)[bl]{applica-}}
\put(13,21){\makebox(0,0)[bl]{tions}}
\put(14,17){\makebox(0,0)[bl]{\(E_{1}(3)\)}}
\put(14,13){\makebox(0,0)[bl]{\(E_{2}(1)\)}}
\put(14,9){\makebox(0,0)[bl]{\(E_{3}(2)\)}}
\put(14,05){\makebox(0,0)[bl]{\(E_{4}(2)\)}}

\put(00,29){\makebox(0,0)[bl]{Server }}
\put(00,25){\makebox(0,0)[bl]{for DBs}}
\put(00,21){\makebox(0,0)[bl]{\(M_{1}(3)\)}}
\put(00,17){\makebox(0,0)[bl]{\(M_{2}(2)\)}}
\put(00,13){\makebox(0,0)[bl]{\(M_{3}(1)\)}}

\end{picture}
\end{center}

%
 Fig. 17 depicts systems solutions at three stages and
 an example
 of the resultant system trajectory:

 \(\alpha = < S^{1}_{2}, \widehat{S}^{1}_{2}, \overline{S}^{1}_{1} > \).

\begin{center}
\begin{picture}(72,25)
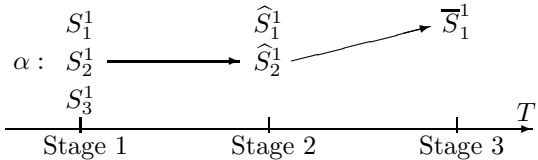
\put(2,0){\makebox(0,0)[bl] {Fig. 17. Illustration for system
trajectory}}

\put(0,8){\vector(1,0){70}} \put(68,9){\makebox(0,8)[bl]{\(T\)}}

\put(5,4){\makebox(0,8)[bl]{Stage 1}}
\put(30,4){\makebox(0,8)[bl]{Stage 2}}
\put(55,4){\makebox(0,8)[bl]{Stage 3}}

\put(10,7.5){\line(0,1){2}} \put(35,7.5){\line(0,1){2}}
\put(60,7.5){\line(0,1){2}}

\put(01,16){\makebox(0,8)[bl]{\(\alpha :\)}}




\put(13.5,17){\vector(1,0){17.6}}

\put(38,17){\vector(4,1){18.5}}

\put(8,20){\makebox(0,8)[bl]{\(S^{1}_{1}\)}}

\put(8,15){\makebox(0,8)[bl]{\(S^{1}_{2}\)}}

\put(8,10){\makebox(0,8)[bl]{\(S^{1}_{3}\)}}


\put(33,20){\makebox(0,8)[bl]{\( \widehat{S}^{1}_{1}\)}}


\put(33,15){\makebox(0,8)[bl]{\( \widehat{S}^{1}_{2}\)}}



\put(58,20){\makebox(0,8)[bl]{\( \overline{S}^{1}_{1}\)}}



\end{picture}
\end{center}

\section{Conclusion and Future Research}

 In the paper,
 a new modular approach
 to compose a configuration of
   applied
  Web-based
  systems
  is suggested.
 The approach is based on
 Hierarchical Morphological Multicriteria Design (HMMD)
 of modular systems and is illustrated by three simplified
 applied examples.
 In HMMD a special lattice-based discrete space is used to evaluate
 quality of the resultant composite system decisions or system configurations.
 The lattice above integrates ordinal quality of system elements and
 ordinal quality of compatibility among the system elements.

 Note, the system structure in HMMD is considered as a tree.
 This is useful from the following viewpoints:
  ~(i) it often allows to construct solving schemes and/or solving algorithms
  with a polynomial complexity;
 more generalized system structures lead to
 NP-hard or/and NP-complete problems;
 ~(ii) tree-like structures are more easy and understandable  for
 readers and end-users,
 and
 it is very important
 to facilitate
 comprehension
 of a new methodology at the 1st steps via simplified structures;
 and
 ~(iii) tree-like structures can be used as a basis for examination
 of more complicated system structures (e.g., hierarchies) and
 approximation of the complicated system structures by tree-like
 structures is an important underlaying approach in solving processes.

 In the future it may be reasonable to consider
 the following research directions:

 {\it 1.} extension of the considered system architecture
  (i.e., examination of hierarchical structures instead of trees);

 {\it 2.} analyzing some issues of system adaptability and
   upgradeability;

 {\it 3.} examination of special new approaches to analysis/comparison
 of the resultant decisions;

%
 {\it 4.} usage of the described lattices of integrated system quality for
 other combinatorial problems which lead to composite solutions
 (e.g., knapsack problem, multiple choice problem);

 {\it 5.} usage of a more complicated lattice-based discrete space of
 system quality that involves poset-like scale for element
 compatibility
 as it was suggested in
 (\cite{levf01}, \cite{lev06});

 {\it 6.} usage of fuzzy set approaches and AI techniques; and

 {\it 7.} examination of other network applications.


%

%
%
%
%
%


\begin{thebibliography}{250}

 \bibitem {aggar00} A. Aggarwal, (Ed.),
  Web-Based Learning and Teaching Technologies: Opportunities and
 Challanges.
 IGI Global, Hershey, PA, 2000.

 \bibitem {arpinar05} I.B. Arpinar, R. Zhang, B. Aleman-Meza, A.
 Maduko,
 Ontology-driven Web services composition platform,
 Information Systems and E-Business Management 2(2) (2005)
  175-199.


 \bibitem {bat01} S. Battacharjee, R. Ramesh, S. Zionts,
  A design framework for e-business infrastructure integration and
  resource management,
  IEEE Trans. on SMC, Part C 31(3) (2001)
   304--319.

 \bibitem {bellazzi01} R. Bellazzi, S. Montani, A. Riva, M.
 Stefanelli,
 Web-based telemedicine systems for home-care: technical issues
 and experiences,
  Computer Methods and Programs in Biomedicine 64(3) (2001)
  175-187.

 \bibitem{benat02} B. Benatallah, M. Dumas, M.-C. Fauvet, F.A.
 Rabhi, Q.Z. Sheng,
  Overview of some patterns for architecting and managing
  composite web services,
   ACM SIGecom Exchanges 3(3) (2002) 9-16.

 \bibitem{benat03} B. Benatallah, Q.Z. Sheng, M. Dumas,
  The self-serv environment for web services composition,
  IEEE Internet Computing 7(6) (2003) 40-48.

%
%



  \bibitem {blake05} M.B. Blake, H. Gomaa,
 Agent-based compositional approaches
 to services-based cross-organizational workflow,
  Decision Support Systems 40(1) (2004) 31-50.


 \bibitem {brus98} P. Brusilovsky, J. Eklund, E. Schwarz,
 Web-based education for all: A tool for development adaptive
 courseware,
  Computer Networks and ISDN Systems 30(1-7) (1998) 291-300.




 \bibitem {cimino95} J.J. Cimino, S.A. Socratous, P.D. Clayton,
 Internet as clinical information system: application development
 using the World Wide Web,
  J. American Medical Informatics Association
 2(5) (1995) 273-284.




%

 \bibitem {dust05} S. Dustdar, W. Schreiner,
 A survey on web services composition,
  Int. J. of Web and Grid Services
  1(1) (2005) 1-30.




   \bibitem {feng06} L. Feng, G. Wang, C. Zeng, R. Huang (Eds.),
  Web Information Systems - WISE 2006 Workshops, LNCS 4256,
 Springer, New York, NY, 2006.


 \bibitem {fra99} P. Fraternali,
 Tools and approaches for developing data-intensive
 Web applications: a survey,
  ACM Computing Surveys  31(3) (1999 227-263.


 \bibitem {gar79} M.R. Garey, D.S. Johnson,
   Computers and Intractability. The
  Guide to the Theory of NP-Completeness,
   W.H.Freeman and Company,
 San Francisco,
  1979.

 \bibitem {geller97} H.-W. Gellersen, R. Wicke, M. Gaedke,
 WebComposition: an object-oriented support system for the Web
 engineering lifecycle,
  Computer Networks and ISDN Systems
 29(8-13) (1997) 1429-1437.

 \bibitem {geng03} X. Geng, Y. Huang, A.B. Winston,
 Smart marketplaces: a step beyind Web services,
 Information Systems and E-Business Management
  1(1) (2003) 15-34.

 \bibitem {ginige01} A. Ginige, S. Murugesan,
 Web engineering: An introduction,
 IEEE Multimedia 8(1) (2001) 14-18.


 \bibitem {heeks05} R. Heeks,
  Implementing and Managing eGovernment.
  Sage Publications, Los Angeles, CA, 2005.

 \bibitem {huang05} Y. Huang, X. Geng, A.B. Winston,
 Network mapping services for dynamic selection of Web services:
 promises and challenges,
  Information Systems and E-Business Management
  3(3) (2005) 281-297.

  \bibitem {insua08} D. R. Insua, G.E. Kersten, J. Rios, C. Crima,
 Towards decision support for participatory democracy,
 Information Systems and E-Business Management
  6(2) (2008) 161-191.

 \bibitem {isa98} T. Isakowitz, M. Bieber, F. Vitali,
 Web information systems,
  Comm. of the ACM 41(7) (1998) 78-80.


  \bibitem {kappel06} G. Kappel, B. Proll, S. Reich, W.
 Retschitzegger, (Eds.),
  Web Engineering - The Discipline of Systematic Development of Web
 Applications.
 J.Wiley \& Sons, New York, NY, 2006.

 \bibitem{kee76} R.L. Keeny, H. Raiffa,
  Decisions with Multiple Objectives:
  Preferences and Value Tradeoffs.
   J.Wiley \& Sons, New York, NY, 1976.

 \bibitem {keller04} H. Kellerer, U. Pferschy, D. Pisinger,
  Knapsack Problems. Springer, Berlin, 2004.

 \bibitem {kingston00} R. Kingston, S. Carver, A. Evans,
  I. Turton,
  Web-based public participation geographical information systems:
  an aid to local environment decision making,
   Computers, Environment, and Urban Systems
   24(2) (2000) 109-125.

 \bibitem{kor84} P. Korhonen, J. Wallenius, and S. Zionts,
  Solving the discrete multiple criteria problems using convex
  cones,
   Manag. Sci. 30(11) (1984) 1336--1345.


 \bibitem {land07} M. Land, B. Fitzgerald,
 Web-based systems design: a study of contemporary practices
 and an explanatory framework based on ``method-in-action'',
 Requirements Engineering
  12(4) (2007) 203-220.

 \bibitem {layne01} K. Layne, J. Lee,
 Developing fully functional E-government: A four stage model,
 Government Information Quaterly
  18(2) (2001) 122-136.

 \bibitem{lev98} M.Sh. Levin,
   Combinatorial Engineering of Decomposable Systems.
  Kluwer Academic Publishers, Dordrecht, 1998.


 \bibitem {levf01} M.Sh. Levin,
  System synthesis with morphological clique problem: fusion of
  subsystem evaluation decisions,
   Information Fusion 2(3) (2001) 225-237.



  \bibitem {lev05} M.Sh. Levin,
 Modular system synthesis: Example for composite packaged software,
  IEEE Trans. on SMC, Part C  35(4) (2005) 544-553.

  \bibitem {lev06} M.Sh. Levin,
   Composite Systems Decisions,
  Springer, New York, 2006.




 \bibitem {lev09} M.Sh. Levin,
 Combinatorial optimization in system configuration design,
  Autom.\&Rem. Control 70(3) (2009) 519-561.


  \bibitem{levshar09} M.Sh. Levin, S.Yu. Sharov,
 Hierarchical morphological composition of Web-hosting system,
  J. of Integrated Design and Process Science
  13(1) (2009) 1-14.

 \bibitem{levs06} M.Sh. Levin, A.V. Safonov,
 Towards modular redesign of networked system.
 In: 2nd Int. Congress on Ultra Modern Telecomm.
 \& Control Syst. and Workshops ICUMT-2010,
 Moscow, (2010) 109-114.

 \bibitem {leymann02} F. Leymann, D. Roller, M. Schmidt,
  Web services and business process management,
  IBM Systems J.  41(2) (2002) 198--211.


 \bibitem {lin08} W.-L. Lin, C.-C. Lo, K.-M. Chao, M. Younas,
 Consumer-centeric QoS-aware selection of web services,
  J. of Computer and System Sciences
  74(2) (2008) 211-231.


 \bibitem {mad06} T. Madhusudan, N. Uttamsingh,
 A declarative approach to composing web services in dynamic
 environments,
 Decision Support Systems 41(2) (2006) 325--357.


 \bibitem {mano05} I. Manolescu, M. Brambilla, S. Ceri, S. Comai,
 P. Fraternali,
 Model-driven design and deployment of service-enabled
 web applications,
  ACM Trans. on Internet Technology
  5(3) (2005 439-479.


 \bibitem {mar90} S. Martello, P. Toth,
   Knapsack Problem: Algorithms and Computer Implementation.
  J.Wiley \& Sons, New York, NY, 1990.

 \bibitem {mccormack97} C. McCormack, D. Jones,
  Building a Web-Based Education System.
 J.Wiley \& Sons, New York, NY, 1997.



 \bibitem {milanovic04} N. Milanovic, M. Malek,
  Current solutions for web service composition,
   IEEE Internet Computing 8(6) (2004) 51-59.

 \bibitem {moon02} M.J. Moon,
 The evolution of E-government among municipalities,
  Public Administration Review 62(4) (2002) 424-433.





 \bibitem{oh05} S.-C. Oh, D. Lee, S.R.T. Kumara,
  A comparative illustration of AI
  planning-based web services composition,
 ACM SIGecom Exchanges 5(5) (2005) 1-10.

 \bibitem {papa06} M.P. Papazoglou, W.-J. van den Heuvel,
 Service-oriented design and development methodology,
  Int. J. Web Engineering and Technology,
 2(4) (2006) 412-442.





  \bibitem{prins06} J.E.J. Prins,
  Designing E-Government. 2nd ed.,
 Kluwer Academic Publishers, Dordrecht, 2006.


 \bibitem {rein02} I. Reinhartz-Berger, D. Dori, S. Katz,
 OPM/Web-object-process methodology for developing Web
 applications,
  Annals of Software Engineering 13(1-4) (2002) 141-161.

 \bibitem {rey04} J. Reynolds,
  The Complete E-Commerce Book:
 Design, Build and Maintain a Successful Web-Based Business.
 2nd ed.,
 CMP Books, Gilroy, CA, 2004.

 \bibitem {roy96} B. Roy,
  Multicriteria Methodology for Decision Aiding.
 Kluwer Academic Publishers, Dordrecht, 1996.

 \bibitem {saa88} T.L. Saaty,
   The Analytic Hierarchy Process.
  MacGraw-Hill, New York, NY, 1988.

 \bibitem {scherlis03} W.L. Scherlis, J. Eisenberg,
 IT research, innovation, and e-government,
  Commun. of the ACM 46(1) (2003) 67-68.

 \bibitem {schewe05} K.-D. Schewe, B. Thalheim, A.
 Binermann-Zdanowitcz, R. Kaschek, T. Kuss, B. Tschiedel,
 A conceptual view of Web-based E-learning,
 Education and Information Technologies
  10(1-2) (2005) 83-110.


  \bibitem {shaw00} M. Shaw, R. Blanning, T. Strader, A.B. Whinston,
 Handbook on Electronic Commerce.
 Springer, New York, NY, 2000.

  \bibitem {simon58} H.A. Simon, A. Newell,
 Heuristic problem solving: The next advance in operations
 research,
 Operations Research  6(1) (1958) 1-10.


  \bibitem{poladian06} J.P. Sousa, V. Poladian, D. Garlan, B. Schmerl, M. Shaw,
   Task-based adaptation for ubiquitous computing,
   IEEE Trans. on SMC - Part C
    36(3) (2006) 328--340.


%



 \bibitem {sung08} Y.-W. Sung, L. Zhou,
 The effect of alliance information on Web service
 composition,
  Information Systems and E-Business Management
  6(4) (2004) 403-417.


 \bibitem {taka97} K. Taka, E. Liang,
 Analysis and design of Web-based information systems,
  Computer Networks and ISDN Systems
 29(8-13) (1997) 1167-1180.

 \bibitem {tang03} H. Tang, Y. Wu, J.T. Yao, G. Wang,
 CUPTRSS: A web-based research support system,
 In: J.T. Yao, P. Lingras (Eds.),
 {\it Proc. of the Workshop on Applications, Products ansd Services of
 Web-based Support Systems WSS03}, Halifax, Canada, pp. 21-28, 2003.







 \bibitem {xiong10} P. Xiong, Y. Fan, M. Zhou,
 A Petri net approach to analysis and composiiton of Web
 services,
  IEEE Trans. SMC - Part A  40(2) (2010) 376-387.

 \bibitem {yao03} Y.Y. Yao,
 A framework for Web-based research support systems,
 in:  Proc. of 23rd Int. Computer Software and
 Application Conf. COMPSAC 2003,
 Dallas, TX, pp. 601-606, 2003.

 \bibitem {yu05} T. Yu, K.-J. Lin,
 Service selection algorithms for Web services
 selection with end-to-end QoS constraints,
  Information Systems and E-Business Management
  3(2) (2005) 103-126.


  \bibitem {younas05} M. Younas, K.-M. Chao, C. Laing,
  Composition of web service in distributed service oriented
  design activities,
   Advanced Engineering Informatics
    19(2) (2005) 143-153.


  \bibitem {zwi69} F. Zwicky,
   Discovery Invention, Research Through the
  Morphological Approach.
  McMillan, New York, 1969.

 \end{thebibliography}
\end{document}